\documentclass[a4paper,11pt]{article}
\usepackage[version=3]{mhchem}
\usepackage{pos}
\usepackage{amsmath,pdftexcmds}
\usepackage{booktabs}
\usepackage{multirow}
\usepackage{tabularx}
\usepackage{amsfonts}
\usepackage{mathtools}
\usepackage{amsbsy}
\usepackage[T1]{fontenc}
\usepackage{dsfont}
\usepackage{hyperref}

\definecolor{DarkMidnightBlue}{rgb}{0.0, 0.04, 0.14}

\renewcommand{\logo}{\relax}

\hypersetup{hidelinks,
backref=true,
pagebackref=true,
hyperindex=true,
breaklinks=true,
colorlinks=true,
linkcolor=blue, % default= red
citecolor=blue, % default= green
urlcolor=blue,
bookmarks=true,
bookmarksopen=false,
pdftitle={Title},
pdfauthor={Author}}

\DeclareFontFamily{U}{cbgreek}{}
\DeclareFontShape{U}{cbgreek}{m}{n}{
        <-6>    grmn0500
        <6-7>   grmn0600
        <7-8>   grmn0700
        <8-9>   grmn0800
        <9-10>  grmn0900
        <10-12> grmn1000
        <12-17> grmn1200
        <17->   grmn1728
      }{}
\DeclareFontShape{U}{cbgreek}{bx}{n}{
        <-6>    grxn0500
        <6-7>   grxn0600
        <7-8>   grxn0700
        <8-9>   grxn0800
        <9-10>  grxn0900
        <10-12> grxn1000
        <12-17> grxn1200
        <17->   grxn1728
      }{}

\DeclareRobustCommand{\digamma}{%
  \text{\usefont{U}{cbgreek}{\normalorbold}{n}\symbol{147}}%
}

\makeatletter
\newcommand{\normalorbold}{%
  \ifnum\pdf@strcmp{\math@version}{bold}=\z@ bx\else m\fi
}
\makeatother

\title{Electromagnetic Selection Rules for \ce{^{24}Mg} in a 6$\alpha$ Cluster Model with $\mathcal{D}_{4h}$ Symmetry}
\author*[a]{Gianluca Stellin}
\author[b]{Karl-Heinz Speidel$^{\dagger,}$\hspace{-2mm}\note[$\dagger$]{Deceased on June 19, 2023.}}

\affiliation[a]{DRF/IRFU/DPhN/LENA, ESNT, CEA Paris-Saclay, \\
91191 Gif-sur-Yvette, France}

\affiliation[b]{Helmholtz Institut f\"ur Strahlen- und Kernphysik, Universit\"at Bonn,\\
Nu\ss{}allee 14-16, 53115 Bonn, Germany}

\emailAdd{gianluca.stellin@cea.fr}
\abstract{
In the framework of a macroscopic $\alpha$-cluster model,
the structural properties and the spectroscopy of the \ce{^{24}Mg} nucleus are investigated. Special attention is devoted to the electromagnetic selection rules imposed by the point-symmetry group $\mathcal{D}_{4h}$ that leaves invariant the adopted $6\alpha$ equilibrium configuration, a square bipyramid. The analysis entails the application of group-theoretical identities and character tables, in a way familiar to quantum chemists. 
The results show that the occurrence of interband E0, E2, and M1, M2, M3 transitions is strictly regulated by the transformation properties of the excited vibrational modes to which the states in the process belong. Unlike the \ce{^{12}C} nucleus in the $\mathcal{D}_{3h}$-symmetric $3\alpha$ arrangement, M1 transition channels are active between states corresponding to a single quantum of vibrational excitation. Conversely, the measured E1 strengths in the \ce{^{24}Mg} spectrum are attributable to the excitation of single-nucleon degrees of freedom, as E1 transitions are forbidden by the model.
The present investigation is a only part of a wider work, encompassing the spectrum and the whole electromagnetic properties of this nucleus in the considered $\mathcal{D}_{4h}$-symmetric configuration, in preparation.
}

\FullConference{
\begin{center}
\begin{minipage}{0.20\columnwidth} 
\includegraphics[width=0.90\columnwidth]{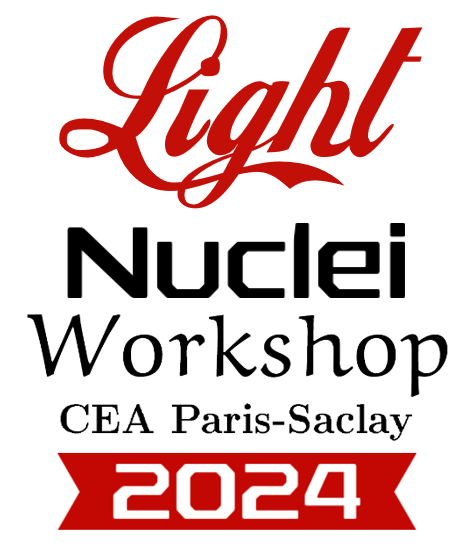} 
\end{minipage}
\begin{minipage}{0.40\columnwidth}
\textbf{ESNT Workshop}\\
\guillemotleft Light nuclei between single- \\
particle and clustering features\guillemotright\\
December 3 - 6, 2024 \\
B\^at. 703, CEA Paris-Saclay,\\
91191 Saint-Aubin, \\
France
\end{minipage}
\end{center}
}

\begin{document}
\maketitle

\section{Introduction}

The recent investigation on the spectrum and electromagnetic properties of \ce{^{20}Ne} \cite{BiI21-01} together with the application of the cluster shell model (CSM) to \ce{^{21}Ne} and \ce{^{21}Na} \cite{BiI21-02} have revived the interest in $\alpha$-clustering on $sd$-shell nuclei. Additionally, the striking reconstruction of the
$3\alpha$ intrinsic shape of \ce{^{12}C} by means of experimental electron-scattering data with minimal theoretical assumptions has enforced the motivation \cite{KiT24}. The latter study provides additional support to the fact that not only the states lying in the Hoyle band exhibit developed $\alpha$-cluster features, but also for the levels in the ground state band the \ce{^4He} clusters represent relevant nuclear subunits. 

Since the inception of nuclear physics \cite{Whe37,Wef37,HaT38}, macroscopic $\alpha$-cluster models, \textit{i.e.} phenomenological approaches retaining the \ce{^4He} clusters as the only degrees of freedom have been applied to $\alpha$-conjugate nuclei, with special attention to \ce{^{12}C} \cite{GlG56,PoC79,BiI02,Jen16,SFV16} and \ce{^{16}O} \cite{Den40,RBL06,BiI14,BiI17,For24}. 

Becoming conglomerates of a small number of $\alpha$-particles, nuclei acquire molecular shapes, differing markedly from the ones of quadrupole (prolate or oblate) or octupole (pear-shaped) type, which originate from the deformation of a continuous spherical surface. The associated finite discrete invariances are referred to as \textit{exotic} nuclear symmetries \cite{DCD18,DeD24}. 
These groups underlie also the macroscopic $\alpha$-cluster approaches developed in relatively recent times, such as the \textit{algebraic cluster model} (ACM) \cite{BiI00,BiI20} and the \textit{quantum graph model} (QGM) \cite{Hal17,Raw18,HaR18}.

Concerning $sd$-shell nuclei, literature adopting macroscopic $\alpha$-particle frameworks is more defective \cite{NTD66,HWD71}. The comprehensive study on intrinsic shapes of \ce{^{20}Ne}, \ce{^{24}Mg}, \ce{^{28}Si}, \ce{^{32}S}, \ce{^{36}Ar} and \ce{^{40}Ca} in Ref.~\cite{HWD71}, needs to be updated, as the significant amount of experimental measurements cumulated since then can potentially jeopardize the predictions. This is the case of intrinsic shape of \ce{^{20}Ne}, predicted as a distorted bitetrahedral configuration of $\alpha$-particles with $\mathcal{D}_{2d}$ symmetry \cite{Hau71,HWD71}, it is now expected to possess a triangular bipyramidal structure with $\mathcal{D}_{3h}$ symmetry, as conjectured in earlier works \cite{HaD66,BFW70}. 

Analogously, although Refs.~\cite{Hau71,HWD71} attributes a bitetrahedral $\alpha$-particle configuration with $\mathcal{D}_{2h}$ symmetry to the \ce{^{24}Mg} nucleus, a very recent study \cite{Ste26} finds more evidence in the spectrum of a square bipyramidal configuration with $\mathcal{D}_{4h}$ symmetry, as indicated by previous literature \cite{HaD66}.
Evidence of a given intrinsic shape can be inferred indirectly from the spin and parity of the measured energy levels as well as from the reduced electric and magnetic multipole transition probabilities. As shown for \ce{^{12}C} \cite{FSV17}, electromagnetic transitions follow patterns which can be constrained by selection rules, stemming from the point group of rotations and reflections which leave a particular nuclear intrinsic shape invariant.

\begin{figure}[ht!]
    \centering
    \begin{minipage}{0.49\columnwidth}
    \includegraphics[width=0.99\columnwidth]{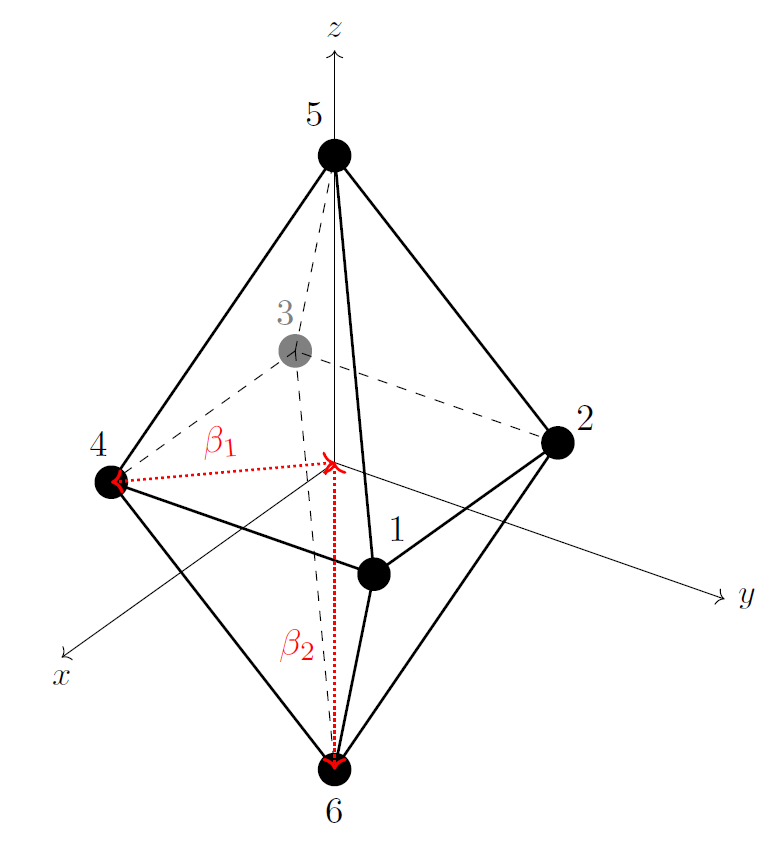}
        \label{fig:MicroscopicD4hStructure}
        \end{minipage}
    \begin{minipage}{0.49\columnwidth}
    \hspace{4mm}
        \includegraphics[width=0.85\columnwidth]{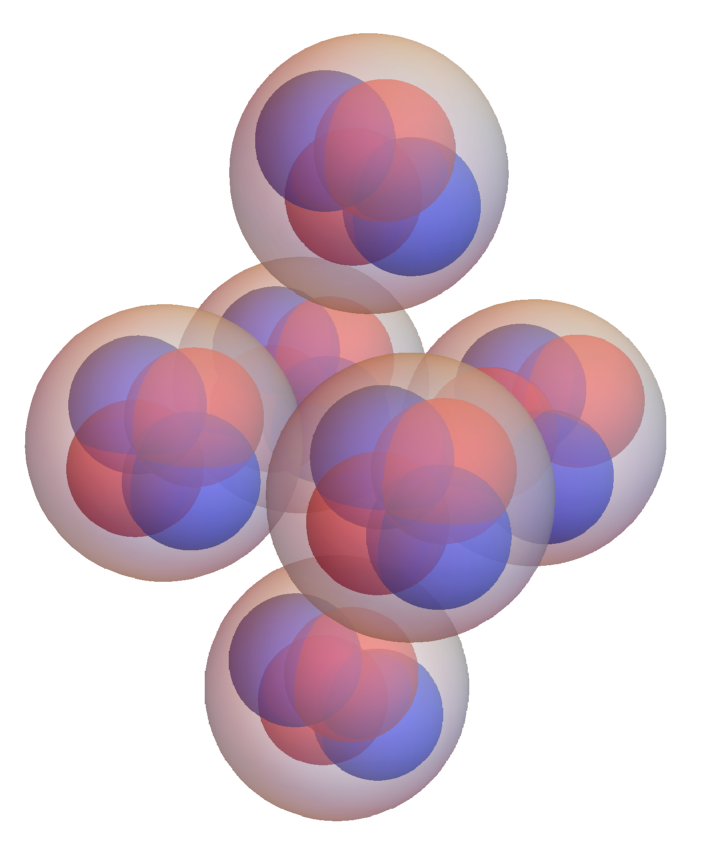}
    \label{fig:MacroscopicD4hStructure}
    \end{minipage}
    \caption{Equilibrium $\alpha$-cluster configuration of \ce{^{24}Mg} in the intrinsic reference frame (left) and with the underlying microscopic structure in terms for protons (red) and neutrons (blue) with realistic charge radii (right). The structure parameters, highlighted in red, are evaluated at (2.38, 3.72) fm, corresponding to a prolate shape, in good agreement with the measured charge radius of the $0_1^+$ state and the electric quadrupole moment of the $2_1^+$ state. The charge distribution of the $\alpha$-particles is assumed to be pointlike.}
    \label{fig:24Mg_SquareBipyramid_D4hStructure}
\end{figure}

\section{The Hamiltonian}

The adopted macroscopic approach coincides with the \textit{geometric $\alpha$-cluster model} (G$\alpha$CM) \cite{Ste26}, according to which the \ce{^{24}Mg} nucleus is described in terms of 6 $\alpha$-particles rotating and vibrating collectively about their equilibrium positions \cite{Ste15}, sitting at the vertices of a \textit{polyhedral} structure, which coincides with a square bipyramid (cf. Fig.~\ref{fig:24Mg_SquareBipyramid_D4hStructure}). 

The most general Hamiltonian of a system of 6 harmonically vibrating and rotating clusters is provided by Watson's Hamiltonian \cite{Wat68}, in which the two collective motions are fully coupled together. However, if the displacements of the $\alpha$-clusters, $\Delta\alpha_i$\footnote{The notation of Refs.~\cite{BuJ04,Ste15} is adopted henceforth: $\alpha,\beta,\gamma \ldots = x,y,z$ ($A,B,C, \ldots =\xi, \eta, \zeta$) are the Cartesian components of tensors defined in the body-fixed or \textit{intrinsic} (laboratory) frame.}, from their equilibrium positions in the body-fixed reference frame, $\alpha_i^e$, are small with respect to the size of the system, characterized by the parameters $(\beta_1,\beta_2)$ in Fig.~\ref{fig:24Mg_SquareBipyramid_D4hStructure}, it is possible to build up the rotation-vibration coupling contributions on top of the rigid-rotor Hamiltonian. In the G$\alpha$CM framework, rotation-vibration coupling terms are obtained from the expansion of the \textit{effective} reciprocal inertia tensor,
\begin{equation}
\mu_{\alpha\beta}^{-1} = I_{\alpha\beta} - \sum_{k=1}^{3N-6}\left( \sum_{j=1}^{3N-6}\zeta_{jk}^{\alpha}Q_j\sum_{l=1}^{3N-6}\zeta_{lk}^{\beta}Q_l\right)~,
\end{equation}
into power series,
\begin{equation}
\boldsymbol{\mu} = (\mathbf{I}^{\mathrm{stat}} + \mathbf{I}^{\mathrm{dyn},\zeta})^{-1} = \left(\mathds{1} - {\mathbf{I}^{\mathrm{stat}}}^{-1}\mathbf{I}^{\mathrm{dyn},\zeta} + {\mathbf{I}^{\mathrm{stat}}}^{-1}\mathbf{I}^{\mathrm{dyn},\zeta}{\mathbf{I}^{\mathrm{stat}}}^{-1}\mathbf{I}^{\mathrm{dyn},\zeta} + \ldots\right){\mathbf{I}^{\mathrm{stat}}}^{-1}\label{eqn:EffectiveInertiaTensorExpansion}
\end{equation}
where $\zeta_{jk}^{\gamma} \in \mathbb{R}$ are c-numbers, defined in Ref.~\cite{Ste15}, $Q_i$ with $i=1,12$ are the \textit{normal coordinates} of vibration, $I_{\alpha\beta}$ is '$\alpha\beta$' component of the inertia tensor, split into the \textit{static} contribution, $I_{\alpha\beta}^{\mathrm{stat}}$, depending solely on the equilibrium $\alpha$-cluster positions, and the \textit{dynamic} contribution, $I_{\alpha\beta}^{\mathrm{dyn}}$, depending quadratically on the normal coordinates. Finally, $I_{\alpha\beta}^{\mathrm{dyn, \zeta}}$
denote the components of the \textit{effective} dynamic inertia tensor,
\begin{equation}
I_{\alpha\beta}^{\mathrm{dyn, \zeta}} \equiv I_{\alpha\beta}^{\mathrm{dyn}} - \sum_{k=1}^{12}\left( \sum_{j=1}^{12}\zeta_{jk}^{\alpha}Q_j\sum_{l=1}^{12}\zeta_{lk}^{\beta}Q_l\right)~.
\end{equation}
The contributions in Eq.~\eqref{eqn:EffectiveInertiaTensorExpansion}, together with the \emph{vibrational} angular momentum, 
\begin{equation}
p_{\alpha}=-i\hbar\sum_{jk=1}^{12}\zeta_{jk}^{\alpha}Q_j\frac{\partial}{\partial Q_k}~,
\end{equation}
are grouped consistently with a power-counting scheme, which permits to obtain a tower of systematically-improved Hamiltonians, $H_{LO}$, $H_{NLO}$, $H_{N^2LO} \ldots$ \cite{Ste26} corresponding to higher-order approximations of Watson's Hamiltonian \cite{Wat68}.
The various interaction terms coupling rotations with vibrations can be treated in perturbation theory, built on top of the reference LO Hamiltonian.

In this context, for the purpose of discussing the EM transition operators and the associated selection rules, it is sufficient to concentrate on the LO Hamiltonian, coinciding with the \textit{rigid-rotor} limit,
\begin{equation}
H_{LO} = \frac{J^2}{2I_{xx}^{\mathrm{stat}}}-\frac{J_z^2}{2}\left(\frac{1}{I_{xx}^{\mathrm{stat}}}-\frac{1}{I_{zz}^{\mathrm{stat}}}\right) - \frac{\hbar^2}{2}\sum_{j=1}^{12}\frac{\partial^2}{\partial Q_j^2} + \frac{1}{2}\sum_{j=1}^{12}\lambda_j^2 Q_j^2 - \frac{\hbar^2}{8}\sum_{\alpha}\mu_{\alpha\alpha}^{\mathrm{stat}}~,
\label{eqn:RigRotHamiltonian}
\end{equation}
where $(\lambda_1,~\lambda_2~\ldots~\lambda_{12})$ $ \equiv (\omega_1,\omega_2,\omega_3, \omega_4, \omega_5, \omega_6,\omega_7,\omega_7,\omega_8,\omega_8,\omega_9,\omega_9)$ are the frequencies associated with the normal modes with coordinates $Q_1,~Q_2,~\ldots~Q_{12}$ respectively
and 
\begin{subequations}
\begin{equation}
I_{xx}^{\mathrm{stat}} = I_{yy}^{\mathrm{stat}} = 2m (\beta_1^2 + \beta_2^2)~, 
\label{eqn:StatInertiaTensor_pl}
\end{equation}
\begin{equation}
I_{zz}^{\mathrm{stat}} = 4m\beta_1^2~,
\label{eqn:StatInertiaTensor_ax}
\end{equation}
\end{subequations}
are the only non-vanishing components of the static inertia tensor, as the axes of the body-fixed frame are parallel to the principal axes of inertia of the square bipyramid and $m \approx 3727.4~\mathrm{MeV}$ is the $\alpha$-particle mass. Besides, the components of the \emph{rotational} angular momentum operator along the body-fixed axes are given by
\begin{subequations}
\begin{equation}
J_x = -i\hbar\Big\{\cos\varphi\left[\cot\theta\frac{\partial}{\partial \varphi} - \frac{1}{\sin\theta}\frac{\partial}{\partial\chi}\right] + \sin\varphi\frac{\partial}{\partial\theta}\Big\}~,
\end{equation}
\begin{equation}
J_y = -i\hbar\Big\{\sin\varphi\left[\cot\theta\frac{\partial}{\partial \varphi} - \frac{1}{\sin\theta}\frac{\partial}{\partial\chi}\right] - \cos\varphi\frac{\partial}{\partial\theta}\Big\}~,
\end{equation}
\begin{equation}
J_z = i\hbar\frac{\partial}{\partial \varphi}~,
\label{eqn:IntrinsicJzeta}
\end{equation}
\end{subequations}
in terms of the Euler angles $(\chi,\theta,\varphi)$ in the active picture \cite{VdW01}. The invariances of $H_{LO}$ are larger than the ones of Watson's Hamiltonian \cite{Wat68}, since rotations and vibrations are decoupled and the system behaves as a symmetric top (cf. Eq.~\eqref{eqn:StatInertiaTensor_pl}-\eqref{eqn:StatInertiaTensor_ax}).  
Since $H_{LO}$ commutes with the components of the angular momentum operator in the laboratory frame, $J_{\xi}$, $J_{\eta}$ and $J_{\zeta}$, as well as with $J_{z}$ in Eq.~\eqref{eqn:IntrinsicJzeta}, the eigenvalues of Eq.\eqref{eqn:RigRotHamiltonian} can be written in close form,
\begin{equation}
E_{LO}(J,K,[\mathfrak{n}]) = \frac{\hbar^2}{2 I_{xx}^{\mathrm{stat}}}[J(J+1) -  K^2] + \frac{\hbar^2}{2 I_{zz}^{\mathrm{stat}}}K^2 + \sum_{i=1}^6\hbar\omega_i(\mathfrak{n}_i+ \frac{1}{2}) + \sum_{i=7}^9\hbar\omega_i(\mathfrak{n}_i+ 1) - \frac{\hbar^2}{8}\sum_{\alpha}{I_{\alpha\alpha}^{\mathrm{stat}}}^{-1}~,
\label{eqn:HLOeigenvalues}
\end{equation}
where $\hbar K$ ($\hbar M$) is the angular momentum projection along the intrinsic (laboratory-fixed) $z$-axis and $\hbar^2 J(J+1)$ is the eigenvalue of the quadratic Casimir operator of $\mathfrak{so}(3)$, $J^2$. In Eq.~\eqref{eqn:HLOeigenvalues}, the number of vibrational quanta for the normal modes $\mathfrak{n}_i$ are vectorized as $[\mathfrak{n}]$, where $i=1,2\ldots 9$, because the modes $i=7,8$ and $9$ are two-dimensional. 

Since rotations and vibrations are decoupled, the eigenstates of $H_{LO}$ can be factorized into a rotational, $\psi_R$, and a vibrational part, $\psi_V$:  
\begin{equation}
\Psi_{RV} \equiv \langle \mathbf{Q}, \boldsymbol{\Omega} \lvert L^{\pi} M, \boldsymbol{\nu} \rangle \equiv \psi_R (\chi,\theta,\varphi)  \psi_V(Q_1, Q_2 \ldots Q_{3N-6})~,
\label{eqn:rovibrationalstatesLO}
\end{equation}
where $\mathbf{Q}$ denotes the normal coordinates in vector form and $\boldsymbol{\Omega}$ the Euler angles. By encoding the vibrational quanta or~\guillemotleft phonons\guillemotright~with the frequencies $\lambda_i$, \textit{i.e.} $\nu_i = \mathfrak{n}_i$ with $i=1,2,\ldots 6$ for the non-degenerate modes and $\nu_7+\nu_8 = \mathfrak{n}_7$, $\nu_9+\nu_{10} = \mathfrak{n}_8$ and $ \nu_{11}+\nu_{12} = \mathfrak{n}_{9}$ for the doubly-degenerate ones, the vibrational eigenfunctions can be succinctly expressed as
\begin{equation}
 \psi_V(Q_1, Q_2 \ldots Q_{3N-6}) =\prod_{i=1}^{12}  \Phi_{{\nu}_i}(Q_i)~,
 \label{eqn:vibrationalstatesLO}
\end{equation}
where $\Phi_{{\nu}_i}$ are one-dimensional harmonic-oscillator eigenfunctions with frequency $\omega_i$,
\begin{equation}
\Phi_{{\nu}_i}(Q_i) = \frac{1}{\sqrt{\nu_i!~2^{\nu_i}}} \left(\frac{\omega_i}{\pi\hbar}\right)^{1/4} H_{\nu_i}({\textstyle\tiny{ \sqrt{\frac{\omega_i}{\hbar}} }}Q_i) e^{-\frac{\omega_i}{2\hbar}Q_i^2}~,
\label{eqn:HOvibreigenfunctions}
\end{equation}
and $H_{\nu_i}$ is a Hermite polynomial of degree $\nu_i$. The HO eigenstates of the doubly-degenerate modes can be equivalently recast into the polar basis \cite{KBH14}, as in Ref.~\cite{Ste15}. 

Since $H_{LO}$ is invariant under parity, $\mathscr{P}$, and time reversal, $\mathscr{T}$, the eigenvectors of the Hamiltonian in Eq.~\ref{eqn:RigRotHamiltonian} transform according to irreducible representations of the two point groups, isomorphic to the cyclic group of order two, $C_2$, in Sch\"onflies notation \cite{Car97}. Due to $\mathscr{P}$ and $\mathscr{T}$ symmetries, the parity and time-reversal operators do not affect the orientation of the body-fixed frame, \textit{i.e.} the Euler angles. 

As a consequence, the parity $\pi$ of the full eigenstate of $H_{LO}$ is determined the one of the vibrational state, $\psi_V$, whereas the rotational wavefunctions, $\psi_R$, possess even parity and are given by
\begin{equation}
\psi_{R}(\chi,\theta,\varphi) \equiv \langle \chi, \theta, \varphi \lvert J, M, K \rangle = \sqrt{\frac{(2J+1)}{8\pi^2}}D_{KM}^{J*}~,
\label{eqn:rotstatesRR}
\end{equation}
where $D_{KM}^{J*}(\chi,\theta,\varphi)$ are Wigner D-matrices in the active picture \cite{VdW01}. In fact, since $SO(3) \supset SO(2)$ constitute dynamical symmetries of $H_{LO}$, referring to 3D rotations in the laboratory-fixed frame and 2D rotations along the z-axis of the intrinsic frame, 
the rotational states can be labeled with the eigenvalues of the Casimir operators of the algebras of the former two groups, $\mathfrak{so}(3)$ and $\mathfrak{so}(2)$ \cite{GrM96}.
However, due to the presence of a combination of an axial symmetry and a point-group symmetry, $\mathcal{D}_{4h}$, the factorization of the eigenstates of $H_{LO}$ into rotational and vibrational states (cf. Eq.~\eqref{eqn:rovibrationalstatesLO}) is no longer valid, in general, for $K \neq 0$ states (cf. Sec.~4-2c of Ref.~\cite{BoM75-II}), as pointed out in the next section.

\section{Normal modes of vibration}

Besides parity and time reversal, another discrete symmetry of the LO G$\alpha$CM Hamiltonian is given by the permutation-inversion group, $\mathcal{G}(6) \equiv \mathcal{S}_6 \times \mathscr{P}$, of the $\alpha$-clusters, which are identical bosons, with zero spin and isospin. As a consequence, the eigenstates of $H_{LO}$ must be completely symmetric under exchange of $\alpha$-clusters. This property holds identically for the eigenstates of the Watson Hamiltonian \cite{BuJ04,Ste26}.

\begin{figure}[ht!]
    \centering
    \includegraphics[width=0.99\columnwidth]{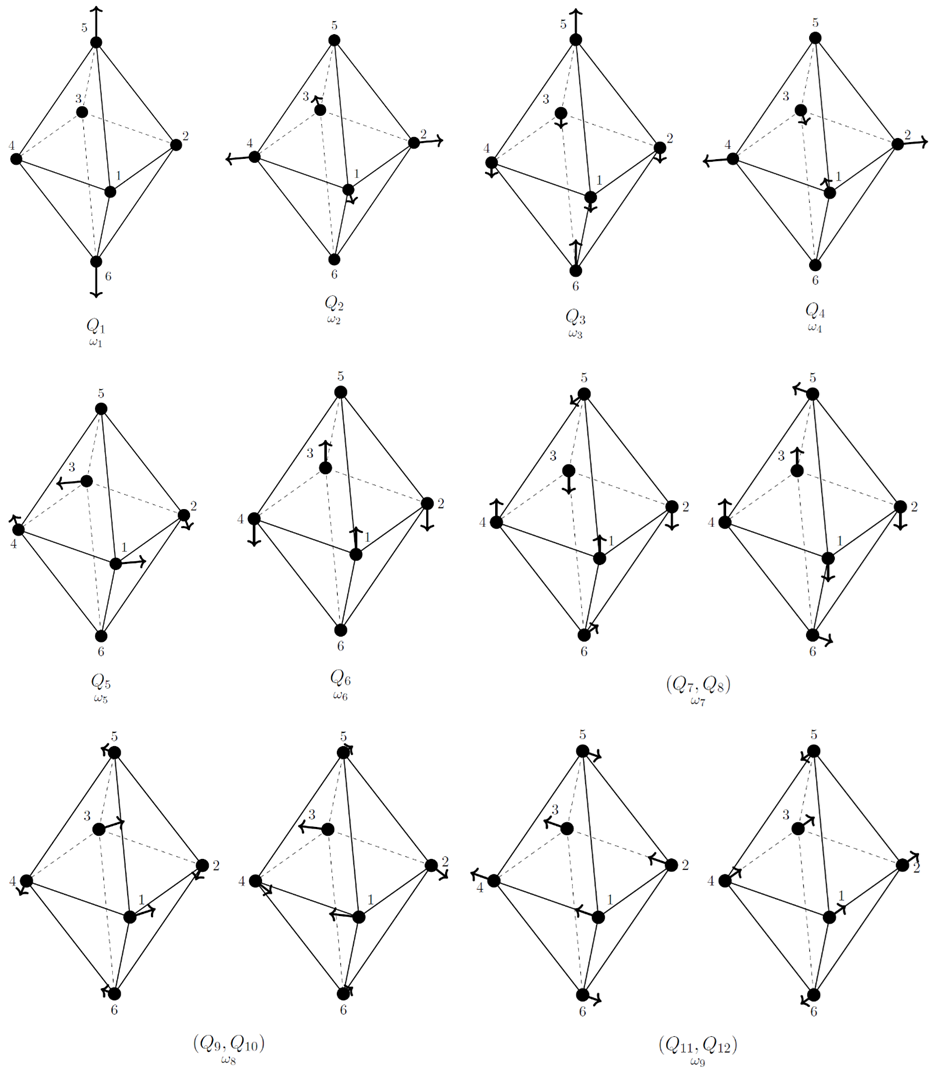}
    \caption{
    Normal vibrations of a square bipyramidal configuration with $\mathcal{D}_{4h}$ symmetry. The oriented segments with arrows denote the displacements of the $\alpha$-clusters with respect to their equilibrium positions.
    }
    \label{fig:24Mg_D4h_NormalModes}
\end{figure}

Nonetheless, the most relevant discrete symmetry for this analysis is represented by the group of operations which leave the equilibrium $\alpha$-cluster arrangement invariant, $\mathcal{D}_{4h}$. Accordingly, the eigenstates of $H_{LO}$, have well-defined transformation properties under the operations of $\mathcal{D}_{4h}$. However, at higher orders in the approximation scheme for the Watson Hamiltonian, triaxiality is induced and the group $\mathcal{D}_{4h}$ no longer represents an exact symmetry of the Hamiltonian \cite{Ste26}.

\begin{table}[hbt!]
\centering
\begin{tabular}{c|ccc ccc cccc|c}
\toprule
 $\mathcal{D}_{4h}$ & $\mathbb{I}$ & $2C_4(z)$ & $C_2$ & $2C'_2$ & $2C''_2$ & $i$ & $2S_4$ & $\sigma_h$ & $2\sigma_v$ & $2\sigma_d$ & \textsc{Coordinates}\\
\midrule
 $A_{1g}$ & $1$ & $1$ & $1$ & $1$ & $1$ & $1$ & $1$ & $1$ & $1$ & $1$ & $Q_1,Q_2$\\
 $A_{2g}$ & $1$ & $1$ & $1$ & $-1$ & $-1$ & $1$ & $1$ & $1$ & $-1$ & $-1$ & \\
 $B_{1g}$ & $1$ & $-1$ & $1$ & $1$ & $-1$ & $1$ & $-1$ & $1$ & $1$ & $-1$ & $Q_4$\\
 $B_{2g}$ & $1$ & $-1$ & $1$ & $-1$ & $1$ & $1$ & $-1$ & $1$ & $-1$ & $1$ & $Q_5$\\
 $E_{g}$ & $2$ & $0$ & $-2$ & $0$ & $0$ & $2$ & $0$ & $-2$ & $0$ & $0$ & $ (Q_7, Q_8)$\\
 $A_{1u}$ & $1$ & $1$ & $1$ & $1$ & $1$ & $-1$ & $-1$ & $-1$ & $-1$ & $-1$ & \\
 $A_{2u}$ & $1$ & $1$ & $1$ & $-1$ & $-1$ & $-1$ & $-1$ & $-1$ & $1$ & $1$ & $Q_3$\\
 $B_{1u}$ & $1$ & $-1$ & $1$ & $1$ & $-1$ & $-1$ & $1$ & $-1$ & $-1$ & $1$ & \\
 $B_{2u}$ & $1$ & $-1$ & $1$ & $-1$ & $1$ & $-1$ & $1$ & $-1$ & $1$ & $-1$ & $Q_6$\\
 \multirow{2}{0.5cm}{\centering{$E_{g}$}} & \multirow{2}{0.5cm}{\centering{$2$}} & \multirow{2}{0.5cm}{\centering{$0$}} & \multirow{2}{0.5cm}{\centering{$-2$}} & \multirow{2}{0.5cm}{\centering{$0$}} & \multirow{2}{0.5cm}{\centering{$0$}} & \multirow{2}{0.5cm}{\centering{$-2$}} & \multirow{2}{0.5cm}{\centering{$0$}} & \multirow{2}{0.5cm}{\centering{$2$}} & \multirow{2}{0.5cm}{\centering{$0$}} & \multirow{2}{0.5cm}{\centering{$0$}} & $ (Q_9, Q_{10}) $\\
   &  &  &  &  &  &  &  &  &  &  & $ (Q_{11}, Q_{12}) $\\
\bottomrule
\end{tabular}
\caption{Character table of the dihedral group $D_{4h}$. The group is composed by 16 operations, including the identity (conjugacy class $\mathbb{I}$), two rotations of angles $\pi/4$ and $3\pi/4$ about the intrinsic $z$-axis ($2C_4(z)$), a rotation by $\pi$ about the z-axis ($C_2$), two rotations of $\pi$ about the axes along to the diagonals of the square ($2C'_2$), two rotations of $\pi$ about the axes parallel to the edges of the square and intersecting the barycentre ($2C''_2$) and the 8 improper rotations obtained by performing the proper rotations before (or after) a spatial inversion (conjugacy classes $i$, $2S_4$, $\sigma_h$, $2\sigma_d$ and $2\sigma_v$ in the respective order). For the irreducible representations (leftmost column), the notation by Mulliken \cite{Car97} is adopted.} \label{tab:D4h_CharacterTable}
\end{table}

The properties of the vibrational states depend on the irreducible representations according to which the creation and annihilation operators built on the normal coordinates transform under the operations of $\mathcal{D}_{4h}$. The latter, in turn, descend from the transformation properties of the normal coordinates, which can be investigated as in Refs.~\cite{BuJ04,Ste15}, by expressing them in terms of the displacement coordinates $\Delta\alpha_i$ in the body-fixed frame. In this form, the normal coordinates of the six non-degenerate modes become \cite{Ste26}

\begin{subequations}
\begin{equation}
Q_1 = \sqrt{m}\left(\frac{\Delta z_5}{\sqrt{2}}-\frac{\Delta z_6}{\sqrt{2}}\right)~,
\label{eqn:normalccordinateQ1}
\end{equation}

\begin{equation}
Q_2 = \sqrt{m}\left(\frac{\Delta x_1}{2\sqrt{2}}-\frac{\Delta x_2}{2\sqrt{2}}-\frac{\Delta x_3}{2\sqrt{2}}+\frac{\Delta x_4}{2\sqrt{2}}+\frac{\Delta y_1}{2\sqrt{2}}+\frac{\Delta y_2}{2\sqrt{2}}-\frac{\Delta y_3}{2\sqrt{2}}-\frac{\Delta y_4}{2\sqrt{2}}\right)~,
\label{eqn:normalccordinateQ2}
\end{equation}

\begin{equation}
Q_3 = \sqrt{m}\left(-\frac{\Delta z_1}{2\sqrt{3}}-\frac{\Delta z_2}{2\sqrt{3}}-\frac{\Delta z_3}{2\sqrt{3}}-\frac{\Delta z_4}{2\sqrt{3}}+\frac{\Delta z_5}{\sqrt{3}}+\frac{\Delta z_6}{\sqrt{3}}\right)~,
\label{eqn:normalccordinateQ3}
\end{equation}

\begin{equation}
Q_4 = \sqrt{m}\left(-\frac{\Delta x_1}{2\sqrt{2}}-\frac{\Delta x_2}{2\sqrt{2}}+\frac{\Delta x_3}{2\sqrt{3}}+\frac{\Delta x_4}{2\sqrt{3}}-\frac{\Delta y_1}{2\sqrt{2}}+\frac{\Delta y_2}{2\sqrt{2}}+\frac{\Delta y_3}{2\sqrt{2}}-\frac{\Delta y_4}{2\sqrt{2}}\right)~,
\label{eqn:normalccordinateQ4}
\end{equation}

\begin{equation}
Q_5 = \sqrt{m}\left(-\frac{\Delta x_1}{2\sqrt{2}}+\frac{\Delta x_2}{2\sqrt{2}}+\frac{\Delta x_3}{2\sqrt{3}}-\frac{\Delta x_4}{2\sqrt{3}}+\frac{\Delta y_1}{2\sqrt{2}}+\frac{\Delta y_2}{2\sqrt{2}}-\frac{\Delta y_3}{2\sqrt{2}}-\frac{\Delta y_4}{2\sqrt{2}}\right)~,
\label{eqn:normalccordinateQ5}
\end{equation}

\begin{equation}
Q_6 = \sqrt{m}\left(\frac{\Delta z_1}{2}-\frac{\Delta z_2}{2}+\frac{\Delta z_3}{2}-\frac{\Delta z_4}{2}\right)~.
\label{eqn:normalccordinateQ6}
\end{equation}
\end{subequations}

In particular, the coordinate $Q_1$ represents a symmetric \textit{stretching} or~\guillemotleft axial breathing\guillemotright~mode, the $Q_2$ a symmetric \textit{stretching} or~\guillemotleft planar breathing\guillemotright~mode, the $Q_3$ a symmetric \textit{wagging} mode, the $Q_4$ an asymmetric \textit{stretching} mode, the $Q_5$ a symmetric \textit{scissoring} mode and the $Q_6$ a symmetric \textit{twisting} mode \cite{Ste26} (cf. Fig.~\ref{fig:24Mg_D4h_NormalModes}).

For the doubly-degenerate modes, their expressions in terms of the displacement coordinates in the intrinsic frame are not unique, as valid normal coordinates can be obtained by taking linear combinations of the coordinates in the respective pairs, $(Q_7,Q_8)$, $(Q_9,Q_{10})$ and $(Q_{11},Q_{12})$, under the constraint of normalization to $\sqrt{m}$ \cite{Ste26}. Their expressions give

\begin{subequations}
\begin{equation}
Q_7 = \frac{\sqrt{m}}{\sqrt{\beta_1^2+\beta_2^2}}\left(\frac{\beta_2 \Delta z_1}{2}-\frac{\beta_2 \Delta z_2}{2}-\frac{\beta_2 \Delta z_3}{2}+\frac{\beta_2 \Delta z_4}{2}+\frac{\beta_1 \Delta x_5}{\sqrt{2}}-\frac{\beta_1 \Delta x_6}{\sqrt{2}}\right)~,
\label{eqn:normalccordinateQ7}
\end{equation}

\begin{equation}
Q_8 = \frac{\sqrt{m}}{\sqrt{\beta_1^2+\beta_2^2}}\left(-\frac{\beta_2 \Delta z_1}{2}-\frac{\beta_2 \Delta z_2}{2}+\frac{\beta_2 \Delta z_3}{2}+\frac{\beta_2 \Delta z_4}{2}-\frac{\beta_1 \Delta y_5}{\sqrt{2}}+\frac{\beta_1 \Delta y_6}{\sqrt{2}}\right)~,
\label{eqn:normalccordinateQ8}
\end{equation}

\begin{equation}
\begin{split}
Q_9 = \sqrt{m}\left(-\frac{9\sqrt{3}}{2\sqrt{326}}\Delta x_1 +\frac{9\sqrt{3}}{2\sqrt{326}}\Delta x_2 -\frac{9\sqrt{3}}{2\sqrt{326}}\Delta x_3 +\frac{9\sqrt{3}}{2\sqrt{326}}\Delta x_4 +\frac{11\Delta y_1}{2\sqrt{978}} \right. \\ \left. +\frac{11\Delta y_2}{2\sqrt{978}}+\frac{11\Delta y_3}{2\sqrt{978}}+\frac{11\Delta y_4}{2\sqrt{978}}-\frac{4\sqrt{2}}{\sqrt{489}}\Delta y_5 -\frac{4\sqrt{2}}{\sqrt{489}}\Delta y_6\right)~,
\end{split}
\label{eqn:normalccordinateQ9}
\end{equation}

\begin{equation}
\begin{split}
Q_{10} = \sqrt{m}\left(\frac{11\Delta x_1}{2\sqrt{978}}+\frac{11\Delta x_2}{2\sqrt{978}}+\frac{11\Delta x_3}{2\sqrt{978}}+\frac{11\Delta x_4}{2\sqrt{978}} -\frac{4\sqrt{2}}{\sqrt{489}}\Delta x_5-\frac{4\sqrt{2}}{\sqrt{489}}\Delta x_6-\frac{9\sqrt{3}}{2\sqrt{326}}\Delta y_1 \right. \\ \left. +\frac{9\sqrt{3}}{2\sqrt{326}}\Delta y_2 -\frac{9\sqrt{3}}{2\sqrt{326}}\Delta y_3 +\frac{9\sqrt{3}}{2\sqrt{326}}\Delta y_4 \right)~,
\label{eqn:normalccordinateQ10}
\end{split}
\end{equation}

\begin{equation}
Q_{11} = \sqrt{m}\left(-\frac{5\Delta y_1}{2\sqrt{33}}
-\frac{5\Delta y_2}{2\sqrt{33}}
-\frac{5\Delta y_3}{2\sqrt{33}}
-\frac{5\Delta y_4}{2\sqrt{33}}
+\frac{2\Delta y_5}{\sqrt{33}}
+\frac{2\Delta y_6}{\sqrt{33}}\right)~,
\label{eqn:normalccordinateQ11}
\end{equation}

\begin{equation}
Q_{12} = \sqrt{m}\left(-\frac{5\Delta x_1}{2\sqrt{33}}
-\frac{5\Delta x_2}{2\sqrt{33}}
-\frac{5\Delta x_3}{2\sqrt{33}}
-\frac{5\Delta x_4}{2\sqrt{33}}
+\frac{2\Delta x_5}{\sqrt{33}}
+\frac{2\Delta x_6}{\sqrt{33}}\right)~.
\label{eqn:normalccordinateQ12}
\end{equation}
\end{subequations}

In Eqs.~\eqref{eqn:normalccordinateQ7}~\eqref{eqn:normalccordinateQ12}, the coordinates $(Q_7,Q_8)$ represent an asymmetric \textit{twisting} mode, the $(Q_9,Q_{10})$ an asymmetric \textit{scissoring} mode and the $(Q_{11},Q_{12})$ an asymmetric \textit{rocking} mode \cite{Ste26} (cf. Fig.~\ref{fig:24Mg_D4h_NormalModes}).

By constructing suitable matrix operators for the $16$ elements of $\mathcal{D}_{4h}$ \cite{Ste26}, for the normal coordinates one obtains the transformation properties recapitulated in Tab.~\ref{tab:D4h_CharacterTable}. The irreducible representations $E_{g}$ and $E_u$ are two-dimensional, whereas all the others are 1-dimensional. In the labels in the leftmost column of Tab.~\ref{tab:D4h_CharacterTable}, the superscript \textit{g} (\textit{u}) denotes an irreducible representation of even (odd) parity.

Concerning the vibrational states, the state corresponding to zero vibration quanta, $\mathfrak{n} = \mathbf{0}$, transforms as the trivial irreducible representation, $A_{1g}$. The harmonic oscillator states with a single phonon in the non-degenerate mode $\omega_i$ transform according to the same irreducible representation as the normal coordinate $Q_i$. Analogously, states with $\mathfrak{n}_7$, $\mathfrak{n}_8$ or $\mathfrak{n}_9 =1$ transform according to the $E_g$, $E_u$ and $E_u$ representations respectively. 

For higher-quanta excitations, the eigenfunctions with an odd number of oscillator quanta in the non-degenerate mode $\omega_i$ behave as the coordinate $Q_i$. If the number of phonons in a non-degenerate mode is even, the corresponding harmonic oscillator eigenfunction transforms as the $A_{1g}$ representation.

 \begin{table}[htb!]
 \centering
 \begin{minipage}{0.49\columnwidth}
 \centering
%\begin{small}
\begin{tabular}{c|c}
\toprule
$E_g$: $\mathfrak{n}_7 = $ & $\Gamma[\Phi_{\nu_7}\Phi_{\nu_8}]^{(\mathfrak{n}_7)}$ \\
\midrule
0 & $A_{1g}$ \\
1 & $E_g$\\
2 & $A_{1g}\oplus B_{1g} \oplus B_{2g}$\\
3 & $2 E_g$\\
4 & 2$A_{1g} \oplus A_{2g} \oplus B_{1g} \oplus B_{2g} $\\
5 & $ 3 E_g$\\
6 & $2 A_{1g} \oplus A_{2g} \oplus 2 B_{1g} \oplus 2B_{2g}$\\
%$\Gamma(\Phi_7)= A_1' \oplus A_2' \oplus E' \oplus E' \oplus E'$ & $8$ & -$1$ & $0$ & $8$ & -$1$ & $0$\\
%$\Gamma(\Phi_8)= A_1' \oplus A_1' \oplus A_2' \oplus E' \oplus E' \oplus E'$ & $9$ & $0$ & $1$ & $9$ & $0$ & $1$\\
%$\Gamma(\Phi_9)= A_1' \oplus A_1' \oplus A_2' \oplus A_2' \oplus E' \oplus E' \oplus E'$ & $10$ & $1$ & $0$ & $10$ & $1$ & $0$\\
%$\Gamma(\Phi_{10})= A_1' \oplus A_1' \oplus A_2' \oplus E' \oplus E' \oplus E' \oplus E'$ & $11$ & -$1$ & $1$ & $11$ & -$1$ & $1$\\
\bottomrule
\end{tabular}
%\end{small}
\end{minipage}
 \begin{minipage}{0.49\columnwidth}
 \centering
%\begin{small}
\begin{tabular}{c|c}
\toprule
$E_u$: $\mathfrak{n}_8,\mathfrak{n}_9 = $ & $\Gamma[\Phi_{\nu_9}\Phi_{\nu_{10}}]^{(\mathfrak{n}_8)}$, $\Gamma[\Phi_{\nu_{11}}\Phi_{\nu_{12}}]^{(\mathfrak{n}_9)}$ \\
\midrule
0 & $A_{1g}$ \\
1 & $E_u$\\
2 & $A_{1g}\oplus B_{1g} \oplus B_{2g}$\\
3 & $2 E_u$\\
4 & 2$A_{1g} \oplus A_{2g} \oplus B_{1g} \oplus B_{2g} $\\
5 & $3 E_u$\\
6 & $2 A_{1g} \oplus A_{2g} \oplus 2 B_{1g} \oplus 2B_{2g}$\\
%$\Gamma(\Phi_7)= A_1' \oplus A_2' \oplus E' \oplus E' \oplus E'$ & $8$ & -$1$ & $0$ & $8$ & -$1$ & $0$\\
%$\Gamma(\Phi_8)= A_1' \oplus A_1' \oplus A_2' \oplus E' \oplus E' \oplus E'$ & $9$ & $0$ & $1$ & $9$ & $0$ & $1$\\
%$\Gamma(\Phi_9)= A_1' \oplus A_1' \oplus A_2' \oplus A_2' \oplus E' \oplus E' \oplus E'$ & $10$ & $1$ & $0$ & $10$ & $1$ & $0$\\
%$\Gamma(\Phi_{10})= A_1' \oplus A_1' \oplus A_2' \oplus E' \oplus E' \oplus E' \oplus E'$ & $11$ & -$1$ & $1$ & $11$ & -$1$ & $1$\\
\bottomrule
\end{tabular}
%\end{small}
\end{minipage}
\caption{Transformation properties under the operations of the $\mathcal{D}_{4h}$ group of the doubly degenerate harmonic oscillator states, associated with the normal modes $\omega_7$, $\omega_8$ and $\omega_9$.}\label{tab:transfprop_D4h_doublydegenerateHOstates}
\end{table}

Conversely, the eigenfunctions corresponding to multiple vibration quanta $\mathfrak{n}_i$ in the doubly-degenerate mode with irreducible representation $\Gamma_2=E_g$ or $E_u$ transform according to a reducible representation, $\Gamma_2^{(\mathfrak{n}_i)}$, whose characters, $\chi^{\Gamma_2^{(\mathfrak{n}_i)}}$, are obtained from the symmetric $\mathfrak{n}_i$-th power of the irreducible representation $\Gamma_2$ associated with the relevant normal coordinate pair \cite{BuJ04,Ste15},  
\begin{equation}
\chi^{\Gamma_2^{(\mathfrak{n}_i)}}[R] = \frac{1}{2}\Big\{\chi^{\Gamma_2}[R]\chi^{\Gamma_2^{(\mathfrak{n}_i-1)}}[R]+\chi^{\Gamma_2}[R^{\mathfrak{n}_i}]\Big\}~,\label{eqn:characterrule_doublydegenerate_irreps}
\end{equation}
where $R$ is an element of $\mathcal{D}_{4h}$. For one-dimensional modes with $\mathfrak{n}_i$ phonons and irreducible representation $\Gamma_1$, Eq.~\eqref{eqn:characterrule_doublydegenerate_irreps} delivers the characters of the representation $\Gamma_1$ ($A_{1g}$) if $\mathfrak{n}_i$ is odd (even).
The application of Eq.~\eqref{eqn:characterrule_doublydegenerate_irreps} yields the results in Tab.~\ref{tab:transfprop_D4h_doublydegenerateHOstates}, for $\mathfrak{n}_i\leq 6$.
Finally, the overall vibrational eigenfunction, $\psi_V$, behaves as the reducible representation obtained by the direct product of the representations according to which the single harmonic oscillator states $\Psi_{\nu_i}$ in Eq.~\eqref{eqn:HOvibreigenfunctions}, transform,
\begin{eqnarray}
    \Gamma_V \equiv \Gamma [\psi_V(Q_1,Q_2\ldots Q_{12})] & = &\Gamma[\Phi_{\nu_1}(Q_1)]^{(\mathfrak{n}_1)} \otimes \Gamma[\Phi_{\nu_2}(Q_2)]^{(\mathfrak{n}_2)} \ldots  \Gamma[\Phi_{\nu_6}(Q_6)]^{(\mathfrak{n}_6)} \nonumber \\ & \otimes &\Gamma[\Phi_{\nu_7}(Q_7)\Phi_{\nu_8}(Q_8)]^{(\mathfrak{n}_7)} \otimes \Gamma[\Phi_{\nu_9}(Q_9)\Phi_{\nu_{10}}(Q_{10})]^{(\mathfrak{n}_8)} \\ & \otimes & \Gamma[\Phi_{\nu_{11}}(Q_{11})\Phi_{\nu_{12}}(Q_{12})]^{(\mathfrak{n}_9)}~. 
\end{eqnarray}

For the rotational states, direct inspection of the transformation properties of $\psi_R$ with different angular momenta as in Ref.~\cite{Ste15}, delivers the results in Tab.~\ref{tab:transprop_rotstates}. For the latter, the fact that the $\alpha$-structure possesses at least an invariance under a rotation of angle $\pi$ about an axis orthogonal to the symmetry axis (\textit{e.g.} the intrinsic $y$-axis) has been exploited (cf. Sec. 4-2c of Ref.~\cite{BoM75-II}). As a consequence, all the operations of $\mathcal{D}_{4h}$ behave on the Wigner D-matrices as proper rotations (cf. Sec. 4-2d of Ref.~\cite{BoM75-II}) and $\psi_R$ has positive parity, regardless of $J$ and $K$. 

Concerning the full eigenstates of $H_{LO}$, they represent a system of identical bosons, hence their transformation properties under the operations of $\mathcal{D}_{4h}$ receive further restrictions. The operations of $\mathcal{D}_{4h}$ that correspond to proper rotations (cf. the conjugacy classes $\mathbb{I}$-$2C_2"$ in Tab.~\ref{tab:D4h_CharacterTable}) entail permutations of the labels of the $\alpha$-clusters (cf. Fig.~\ref{fig:24Mg_SquareBipyramid_D4hStructure}). Conversely, the improper rotations (cf. the conjugacy classes $i$-$2\sigma_d$ in Tab.~\ref{tab:D4h_CharacterTable}) imply both a permutation of the labels of the clusters and the transformation of the right-handed body-fixed Cartesian axes into left-handed ones, due to spatial inversion. Consequently, $\Psi_{RV}$ must transform as a one-dimensional representation completely symmetric (symmetric or antisymmetric) under proper (improper) rotations. Therefore, the irreducible representations of $\mathcal{D}_{4h}$ fulfilling these properties are the $A_{1g}$ and the $A_{1u}$ in Tab.~\ref{tab:D4h_CharacterTable}, associated with even and odd parity states respectively \cite{Ste26}. 

\begin{table}[htb!]
\begin{minipage}{0.33\columnwidth}
\centering
\begin{tabular}{c|c|c}
\toprule
$J$ & $|K|$ & $\Gamma[\psi_R]$ \\
\midrule
$0$ & $0$ & $A_{1g}$\\
\midrule
$1$ & $0$ &  $A_{2g}$\\
& $1$ &  $E_{g}$\\
\midrule
$2$ & $0$ & $A_{1g}$\\
& $1$ & $E_g$\\
& $2$ & $B_{1g} \oplus B_{2g}$ \\
\midrule
$3$ & $0$ & $A_{2g}$ \\
& $1$ & $E_g$\\
& $2$ & $B_{1g}\oplus B_{2g}$\\
& $3$ & $E_g$\\
\midrule
$4$ & $0$ & $A_{1g}$ \\
& $1$ & $E_g$\\
& $2$ & $B_{1g}\oplus B_{2g}$\\
& $3$ & $E_g$\\
& $4$ & $A_{1g}\oplus A_{2g}$\\
\midrule
$5$ & $0$ & $A_{2g}$ \\
& $1$ & $E_g$\\
& $2$ & $B_{1g}\oplus B_{2g}$\\
\bottomrule
\end{tabular}
\end{minipage}
\begin{minipage}{0.33\columnwidth}
\centering
\begin{tabular}{c|c|c}
\toprule
$J^{\pi}$ & $|K|$ & $\Gamma[\psi_R]$ \\
\midrule
$5$ & $3$ & $E_g$\\
& $4$ & $A_{1g}\oplus A_{2g}$\\
& $5$ & $E_g$\\
\midrule
$6$ & $0$ & $A_{1g}$ \\
& $1$ & $E_g$\\
& $2$ & $B_{1g}\oplus B_{2g}$\\
& $3$ & $E_g$\\
& $4$ & $A_{1g}\oplus A_{2g}$\\
& $5$ & $E_g$\\
& $6$ & $B_{1g}\oplus B_{2g}$\\
\midrule
$7$ & $0$ & $A_{2g}$ \\
& $1$ & $E_g$\\
& $2$ & $B_{1g}\oplus B_{2g}$\\
& $3$ & $E_g$\\
& $4$ & $A_{1g}\oplus A_{2g}$\\
& $5$ & $E_g$\\
& $6$ & $B_{1g}\oplus B_{2g}$\\
& $7$ & $E_g$\\
\bottomrule
\end{tabular}
\end{minipage}
\begin{minipage}{0.33\columnwidth}
\centering
\begin{tabular}{c|c|c}
\toprule
$J^{\pi}$ & $|K|$ & $\Gamma[\psi_R]$ \\
\midrule
$8$ & $0$ & $A_{1g}$ \\
& $1$ & $E_g$\\
& $2$ & $B_{1g}\oplus B_{2g}$\\
& $3$ & $E_g$\\
& $4$ & $A_{1g}\oplus A_{2g}$\\
& $5$ & $E_g$\\
& $6$ & $B_{1g}\oplus B_{2g}$\\
& $7$ & $E_g$\\
& $8$ & $A_{1g}\oplus A_{2g}$\\
\midrule
$9$ & $0$ & $A_{2g}$ \\
& $1$ & $E_g$\\
& $2$ & $B_{1g}\oplus B_{2g}$\\
& $3$ & $E_g$\\
& $4$ & $A_{1g}\oplus A_{2g}$\\
& $5$ & $E_g$\\
& $6$ & $B_{1g}\oplus B_{2g}$\\
& $7$ & $E_g$\\
& $8$ & $A_{1g}\oplus A_{2g}$\\
& $9$ & $E_g$\\
\bottomrule
\end{tabular}
\end{minipage}
\caption{Transformation properties under the operations of the $\mathcal{D}_{4h}$ group of the rotational states, $\psi_R$, proportional to Wigner D-matrices.}
\label{tab:transprop_rotstates}
\end{table}

In summary, for the construction of rotational-vibrational states with defined transformation properties under $\mathcal{D}_{4h}$ and bosonic symmetry, the following selection rule proves to hold
\begin{equation}
\Gamma[\psi_V] \otimes \Gamma[\psi_R] \supset A_{1g} \vee A_{1u}~,
\label{eqn:selrule_RotationVibrStates}
\end{equation}
meaning that in the decomposition of the reducible representation given by the direct product of the representations of $\psi_V$ and $\psi_R$ the irreducible representations $A_{1g}$ or $A_{1u}$ must appear. The application of Eq.~\eqref{eqn:selrule_RotationVibrStates} to the representations of the vibrational (cf. Tab~\ref{tab:transfprop_D4h_doublydegenerateHOstates}) and rotational states (cf. Tab.~\ref{tab:transprop_rotstates}), permits to obtain the results outlined in Tabs.~\ref{tab:AllowedRVStates_J04} and \ref{tab:AllowedRVStates_J57} for $J \leq 7$ states. Higher-$J$ states are detailed in Ref.~\cite{Ste26}.

In the first row of Tabs.~\ref{tab:AllowedRVStates_J04} and \ref{tab:AllowedRVStates_J57}, the irreducible representations labeling the normal modes with one quantum of vibration are reported, whereas the ~\guillemotleft $1$'s\guillemotright~in the cells below the second row and on the left of the second column indicates that a single rotational-vibrational state transforming as the $A_{1g}$ or $A_{1u}$ representation can be constructed.  

The results in Tabs.~\ref{tab:AllowedRVStates_J04} and \ref{tab:AllowedRVStates_J57} agree with Refs.~\cite{HaD66,HWD71} and show that the number of vibrational quanta constrains the parity, the value of $|K|$ and the angular momentum $J$ of the states. In terms of the eigenstates, the restrictions imposed by $\mathcal{D}_{4h}$ symmetry result into a non-factorized
expression of the eigenfunction of $H_{LO}$,
\begin{equation}
\begin{split}
\Psi_{RV}^{\mathcal{D}_{4h}}(\chi,\theta,\varphi)  = \sqrt{\frac{(2J+1)}{32\pi^2 \Delta(\mathfrak{n},K)}} \Big\{ \left[\psi_V + (-i)^K \bar{\psi}_V\right] D_{-KM}^{J*}(\chi,\theta,\varphi) \\ +(-1)^{J+K+\nu_3}\left[\psi_V + i^K \bar{\psi}_V\right] & D_{KM}^{J*}(\chi,\theta,\varphi)\Big\}~,
\label{eqn:rovibrationalstates_LO_D4h}
\end{split}
\end{equation}
where $\Delta(\mathfrak{n},K) \equiv (\delta_{K0}+1)(1+\delta_{\nu_7\nu_8}\delta_{\nu_9\nu_{10}}\delta_{\nu_{11}\nu_{12}})$ is a prefactor. The two terms within the square brackets on the r.h.s. of Eq.~\ref{eqn:rovibrationalstates_LO_D4h} are mapped one another by time reversal, $\mathscr{T}$, and the~\guillemotleft partner\guillemotright~ of the vibrational state in Eq.~\eqref{eqn:vibrationalstatesLO}, $\bar{\psi}_V$, has been introduced as in Sec.~4-2c of Ref.~\cite{BoM75-II},
\begin{equation}
\bar{\psi}_V = (-1)^{\sum_{i=4}^{12}\nu_i} \left[\prod_{j=1}^{6}  \Phi_{\nu_j}(Q_j)\right] \Phi_{\nu_8}(Q_7) \Phi_{\nu_7}(Q_8) \Phi_{\nu_{10}}(Q_9) \Phi_{\nu_{9}}(Q_{10}) \Phi_{\nu_{12}}(Q_{11}) \Phi_{\nu_{11}}(Q_{12})~.
\label{eqn:conjugate_vibrationalstatesLO}
\end{equation}
On the r.h.s. of Eq.~\eqref{eqn:conjugate_vibrationalstatesLO}, the phonons assigned to the 1D harmonic oscillator wavefunctions of the doubly-degenerate modes are pairwise flipped with respect to their original positions in Eq.~\eqref{eqn:vibrationalstatesLO}, \textit{i.e.} $\nu_7 \leftrightarrow \nu_8$, $\nu_{9} \leftrightarrow \nu_{10}$ and $\nu_{11} \leftrightarrow \nu_{12}$. Furthermore, if $\nu_{i}=\nu_{i+1}$ in all the doubly-degenerate modes, the states in Eq.~\eqref{eqn:rovibrationalstates_LO_D4h} reacquire the factorized form between the vibrational and the rotational part as in Eq.~\eqref{eqn:rovibrationalstatesLO}, even when $K \neq 0$.

In the G$\alpha$CM for \ce{^{24}Mg} \cite{Ste26}, the observed $J^{\pi}$ energy states are grouped into rotational bands, characterized by $|K|^{\pi}$, the number of vibrational quanta $[\mathfrak{n}]$ in the vibrational part of the assigned $\Psi_{RV}^{\mathcal{D}_{4h}}$ states and (a direct sum of) irreducible representations of $\mathcal{D}_{4h}$ associated with the corresponding excited normal mode(s). Candidates for all the singly-excited vibrational bands have been detected \cite{Ste26}, and their composition is mostly coherent with the literature \cite{GFM78,FGH79,CLW93,KYI12,FKA24}. 

Furthermore, the only measured $0_1^-$ state at $12.385(1)~\mathrm{MeV}$ cannot belong to a singly-excited rotational band. In fact, $J^{\pi}=0^-$ states must have at least two excitation quanta. Specifically, considered the fact that $\hbar\omega_7 + \hbar\omega_9 = 12.29(3)~\mathrm{MeV}$ \cite{Ste26} and that $n_7$ and $n_9$ = 1 states transform as the $E_g \otimes E_u= A_{1u} \oplus A_{2u} \oplus B_{1u} \oplus B_{2u}$ representation, the measured level could constitute the head of a $K^{\pi}=0^-$ band with 
$J^{\pi}=$ $0^-$, $1^-$, $2^-$, $3^-$, $\ldots$ states as members. 

More in general, in the absence of internal degrees of freedom such as the vibrational modes, $J^{\pi} = 0^-$ states cannot be constructed in terms of Wigner D-matrices alone. In such cases, one can resort to basis states of irreducible representations of the group of 3D rotations and reflections, O(3), outlined in Refs.~\cite{Won67,Won69}.

Finally, another return of the G$\alpha$CM \cite{Ste26} is that, in the large-amplitude-vibration limit, the $A_{1g}$ ($\omega_2$), $A_{2u}$ ($\omega_3$) and $E_g$ ($\omega_7$) normal modes favour the $2\alpha$+\ce{^{16}O}, $\alpha$+\ce{^{20}Ne} and \ce{^{12}C}+\ce{^{12}C} cluster configurations and decay channels respectively \cite{CRJ23}.

\begin{table}[htb!]
\centering
\begin{tabular}{c|c|cccccccccc}
\toprule
\multicolumn{2}{c|}{$\mathcal{D}_{4h}$} & $A_{1g}$ & $A_{1g}$ & $A_{1g}$ & $A_{2u}$ & $B_{1g}$ & $B_{2g}$ & $B_{2u}$ & $E_{g}$ & $E_{u}$ & $E_{u}$\\
\midrule
\multicolumn{2}{c|}{$[\mathfrak{n}]=0$, except:} & $\emptyset$  & $\mathfrak{n}_1$ & $\mathfrak{n}_2$ & $\mathfrak{n}_3$ & $\mathfrak{n}_4$ & $\mathfrak{n}_5$ & $\mathfrak{n}_6$ & $\mathfrak{n}_7$ & $\mathfrak{n}_8$ & $\mathfrak{n}_9$\\
\midrule
$J^{\pi}$ & $|K|$ & 0 & 1 & 1 & 1 & 1 & 1 & 1 & 1 & 1 & 1\\
\midrule
 $0^+$ & 0 & 1 & 1 & 1 & 0 & 0 & 0 & 0 & 0 & 0 & 0\\
 \midrule
 $0^-$ & 0 & 0 & 0 & 0 & 0 & 0 & 0 & 0 & 0 & 0 & 0\\
  \midrule
 $1^+$ & 0 & 0 & 0 & 0 & 0 & 0 & 0 & 0 & 0 & 0 & 0\\
   & 1 & 0 & 0 & 0 & 0 & 0 & 0 & 0 & 1 & 0 & 0\\
\midrule
 $1^-$ & 0 & 0 & 0 & 0 & 1 & 0 & 0 & 0 & 0 & 0 & 0\\
 & 1 & 0 & 0 & 0 & 0 & 0 & 0 & 0 & 0 & 1 & 1\\
\midrule
 $2^+$ & 0 & 1 & 1 & 1 & 0 & 0 & 0 & 0 & 0 & 0 & 0\\
   & 1 & 0 & 0 & 0 & 0 & 0 & 0 & 0 & 1 & 0 & 0\\
   & 2 & 0 & 0 & 0 & 0 & 1 & 1 & 0 & 0 & 0 & 0\\
\midrule
 $2^-$ & 0 & 0 & 0 & 0 & 0 & 0 & 0 & 0 & 0 & 0 & 0\\
   & 1 & 0 & 0 & 0 & 0 & 0 & 0 & 0 & 0 & 1 & 1\\
   & 2 & 0 & 0 & 0 & 0 & 0 & 0 & 1 & 0 & 0 & 0\\
\midrule
 $3^+$ & 0 & 0 & 0 & 0 & 0 & 0 & 0 & 0 & 0 & 0 & 0\\
   & 1 & 0 & 0 & 0 & 0 & 0 & 0 & 0 & 1 & 0 & 0\\
   & 2 & 0 & 0 & 0 & 0 & 1 & 1 & 0 & 0 & 0 & 0\\
   & 3 & 0 & 0 & 0 & 0 & 0 & 0 & 0 & 1 & 0 & 0\\
\midrule
 $3^-$ & 0 & 0 & 0 & 0 & 1 & 0 & 0 & 0 & 0 & 0 & 0\\
   & 1 & 0 & 0 & 0 & 0 & 0 & 0 & 0 & 0 & 1 & 1\\
   & 2 & 0 & 0 & 0 & 0 & 0 & 0 & 1 & 0 & 0 & 0\\
   & 3 & 0 & 0 & 0 & 0 & 0 & 0 & 0 & 0 & 1 & 1\\
\midrule
 $4^+$ & 0 & 1 & 1 & 1 & 0 & 0 & 0 & 0 & 0 & 0 & 0\\
   & 1 & 0 & 0 & 0 & 0 & 0 & 0 & 0 & 1 & 0 & 0\\
   & 2 & 0 & 0 & 0 & 0 & 1 & 1 & 0 & 0 & 0 & 0\\
   & 3 & 0 & 0 & 0 & 0 & 0 & 0 & 0 & 1 & 0 & 0\\
   & 4 & 1 & 1 & 1 & 0 & 0 & 0 & 0 & 0 & 0 & 0\\
   \midrule
 $4^-$ & 0 & 0 & 0 & 0 & 0 & 0 & 0 & 0 & 0 & 0 & 0\\
   & 1 & 0 & 0 & 0 & 0 & 0 & 0 & 0 & 0 & 1 & 1\\
   & 2 & 0 & 0 & 0 & 0 & 0 & 0 & 1 & 0 & 0 & 0\\
   & 3 & 0 & 0 & 0 & 0 & 0 & 0 & 0 & 0 & 1 & 1\\
   & 4 & 0 & 0 & 0 & 1 & 0 & 0 & 0 & 0 & 0 & 0\\
\midrule
\end{tabular}
\caption{Allowed rotational-vibrational states for six bosons at the vertexes of a square bipyramid, up to one excitation quantum and angular momentum $0 \leq J \leq 4$ and both parities.}
\label{tab:AllowedRVStates_J04}
\end{table}
\clearpage

\begin{table}[thb!]
\centering
\begin{tabular}{c|c|cccccccccc}
\toprule
\multicolumn{2}{c|}{$\mathcal{D}_{4h}$} & $A_{1g}$ & $A_{1g}$ & $A_{1g}$ & $A_{2u}$ & $B_{1g}$ & $B_{2g}$ & $B_{2u}$ & $E_{g}$ & $E_{u}$ & $E_{u}$\\
\midrule
\multicolumn{2}{c|}{$[\mathfrak{n}]=0$, except:} & $\emptyset$  & $\mathfrak{n}_1$ & $\mathfrak{n}_2$ & $\mathfrak{n}_3$ & $\mathfrak{n}_4$ & $\mathfrak{n}_5$ & $\mathfrak{n}_6$ & $\mathfrak{n}_7$ & $\mathfrak{n}_8$ & $\mathfrak{n}_9$\\
\midrule
$J^{\pi}$ & $|K|$ & 0 & 1 & 1 & 1 & 1 & 1 & 1 & 1 & 1 & 1\\
   \midrule
 $5^+$ & 0 & 0 & 0 & 0 & 0 & 0 & 0 & 0 & 0 & 0 & 0\\
   & 1 & 0 & 0 & 0 & 0 & 0 & 0 & 0 & 1 & 0 & 0\\
   & 2 & 0 & 0 & 0 & 0 & 1 & 1 & 0 & 0 & 0 & 0\\
   & 3 & 0 & 0 & 0 & 0 & 0 & 0 & 0 & 1 & 0 & 0\\
   & 4 & 1 & 1 & 1 & 0 & 0 & 0 & 0 & 0 & 0 & 0\\
   & 5 & 0 & 0 & 0 & 0 & 0 & 0 & 0 & 1 & 0 & 0\\
\midrule
 $5^-$ & 0 & 0 & 0 & 0 & 1 & 0 & 0 & 0 & 0 & 0 & 0\\
   & 1 & 0 & 0 & 0 & 0 & 0 & 0 & 0 & 0 & 1 & 1\\
   & 2 & 0 & 0 & 0 & 0 & 0 & 0 & 1 & 0 & 0 & 0\\
   & 3 & 0 & 0 & 0 & 0 & 0 & 0 & 0 & 0 & 1 & 1\\
   & 4 & 0 & 0 & 0 & 1 & 0 & 0 & 0 & 0 & 0 & 0\\
   & 5 & 0 & 0 & 0 & 0 & 0 & 0 & 0 & 0 & 1 & 1\\
\midrule
 $6^+$ & 0 & 1 & 1 & 1 & 0 & 0 & 0 & 0 & 0 & 0 & 0\\
   & 1 & 0 & 0 & 0 & 0 & 0 & 0 & 0 & 1 & 0 & 0\\
   & 2 & 0 & 0 & 0 & 0 & 1 & 1 & 0 & 0 & 0 & 0\\
   & 3 & 0 & 0 & 0 & 0 & 0 & 0 & 0 & 1 & 0 & 0\\
   & 4 & 1 & 1 & 1 & 0 & 0 & 0 & 0 & 0 & 0 & 0\\
   & 5 & 0 & 0 & 0 & 0 & 0 & 0 & 0 & 1 & 0 & 0\\
   & 6 & 0 & 0 & 0 & 0 & 1 & 1 & 0 & 0 & 0 & 0\\
\midrule
 $6^-$ & 0 & 0 & 0 & 0 & 0 & 0 & 0 & 0 & 0 & 0 & 0\\
   & 1 & 0 & 0 & 0 & 0 & 0 & 0 & 0 & 0 & 1 & 1\\
   & 2 & 0 & 0 & 0 & 0 & 0 & 0 & 1 & 0 & 0 & 0\\
   & 3 & 0 & 0 & 0 & 0 & 0 & 0 & 0 & 0 & 1 & 1\\
   & 4 & 0 & 0 & 0 & 1 & 0 & 0 & 0 & 0 & 0 & 0\\
   & 5 & 0 & 0 & 0 & 0 & 0 & 0 & 0 & 0 & 1 & 1\\
   & 6 & 0 & 0 & 0 & 0 & 0 & 0 & 1 & 0 & 0 & 0\\
\midrule
 $7^+$ & 0 & 0 & 0 & 0 & 0 & 0 & 0 & 0 & 0 & 0 & 0\\
   & 1 & 0 & 0 & 0 & 0 & 0 & 0 & 0 & 1 & 0 & 0\\
   & 2 & 0 & 0 & 0 & 0 & 1 & 1 & 0 & 0 & 0 & 0\\
   & 3 & 0 & 0 & 0 & 0 & 0 & 0 & 0 & 1 & 0 & 0\\
   & 4 & 1 & 1 & 1 & 0 & 0 & 0 & 0 & 0 & 0 & 0\\
   & 5 & 0 & 0 & 0 & 0 & 0 & 0 & 0 & 1 & 0 & 0\\
   & 6 & 0 & 0 & 0 & 0 & 1 & 1 & 0 & 0 & 0 & 0\\
   & 7 & 0 & 0 & 0 & 0 & 0 & 0 & 0 & 1 & 0 & 0\\
\midrule
\end{tabular}
\caption{Allowed rotational-vibrational states for six bosons at the vertexes of a square bipyramid, up to one excitation quantum and angular momentum $5 \leq J \leq 7$.}
\label{tab:AllowedRVStates_J57}
\end{table}

\clearpage

\section{Selection Rules}

Let us consider the transitions of electromagnetic type between the various energy states lying in the rotational bands of \ce{^{24}Mg}. Although of different nature, electric monopole transitions are often included in the set.

The reduced electric ($F = E$) or magnetic ($F = M$) multipole transition probability between the initial state $(\Psi_{RV}^{\mathcal{D}_{4h}})_i$ characterized by the parity $\pi_i$, the quantum numbers $J_i$, $M_i$, $K_i$ and $[\mathfrak{n}]_i$ phonons, and the final state $(\Psi_{RV}^{\mathcal{D}_{4h}})_f$ with parity $\pi_f$ and the quantum numbers $J_f$, $M_f$, $K_f$ and $[\mathfrak{n}]_f$, is provided by
\begin{equation}
\begin{split}
B(F\lambda, J_i^{\pi_i}, |K_i|, [\mathfrak{n}]_i \rightarrow J_f^{\pi_f}, |K_f|, [\mathfrak{n}]_f)
 \\ = \frac{1}{2J_i+1}\sum_{Mi=-J_i}^{J_i}\sum_{Mf=-J_f}^{J_f} \sum_{\mu = -\lambda}^{\lambda}  |\langle & J_f, M_f, |K_f|, [\mathfrak{n}]_f |   \Omega_{\lambda\mu}(F)| J_i, M_i, |K_i|, [\mathfrak{n}]_i  \rangle |^2~,
 \end{split}
 \label{eqn:reducedEMtransitionprobabibilites}
\end{equation}
where $\Omega_{\lambda\mu}(F)$ is the transition operator in the laboratory frame, connected with the intrinsic counterpart, $\omega_{\lambda\mu}(F)$, by means of a rotation, 
\begin{equation}
\Omega_{\lambda\mu}(F) = \sum_{\nu = - \lambda}^{\lambda} D_{\mu\nu}^{\lambda}(\varphi,\theta,\chi)~\omega_{\lambda\mu}(F)~.
\end{equation}
where $D_{\mu\nu}^{\lambda}(\varphi,\theta,\chi)= (-1)^{\mu-\nu} D_{\nu\mu}^{\lambda *}(\chi,\theta,\varphi)$ for Wigner D-matrix elements in the active picture \cite{VdW01}. For electric monopole transitions, triggered by the internal conversion mechanism \cite{KGW22}, the operator $\Omega_{00}(E)$ in the laboratory frame is given by 
\begin{equation}
    \Omega_{00}(E) = 2e\sum_{i=1}^{6}\frac{\xi_i^2 + \eta_i^2 + z_i^2}{\langle r^2 \rangle _{g.s.}}~,
\end{equation}
where $\langle r^2 \rangle _{g.s.}$ is the experimental squared charge radius of the $0^+$ ground state of the nucleus. 
\begin{table}[thb!]
\begin{minipage}{0.495\columnwidth}
%\begin{small}
\centering
\begin{tabular}{c|c}
\toprule
$\lambda$ & $\Gamma[\omega_{\lambda\mu}(E)]$\\
\midrule
$0$ & $A_{1g}$\\
$1$ & $A_{2u}\oplus E_u$\\
$2$ & $A_{1g}\oplus B_{1g} \oplus B_{2g} \oplus E_g$\\
$3$ & $A_{2u} \oplus B_{1u} \oplus B_{2u} \oplus 2E_u$\\
$4$ & $2A_{1g} \oplus A_{2g} \oplus B_{1g} \oplus B_{2g} \oplus 2E_g$\\
$5$ & $A_{1u} \oplus 2A_{2u} \oplus B_{1u} \oplus B_{2u} \oplus 3E_u$\\
$6$ & $2A_{1g} \oplus A_{2g} \oplus 2B_{1g} \oplus 2 B_{2g} \oplus 3 E_g$\\
\bottomrule
\end{tabular}
%\end{small}
\end{minipage}
\begin{minipage}{0.495\columnwidth}
%\begin{small}
\centering
\begin{tabular}{c|c}
\toprule
$\lambda$ & $\Gamma[\omega_{\lambda\mu}(M)]$\\
\midrule
$1$ & $A_{2g}\oplus E_g$\\
$2$ & $A_{1u}\oplus B_{1u} \oplus B_{2u} \oplus E_u$\\
$3$ & $A_{2g} \oplus B_{1g} \oplus B_{2g} \oplus 2E_g$\\
$4$ & $2A_{1u} \oplus A_{2u} \oplus B_{1u} \oplus B_{2u} \oplus 2E_u$\\
$5$ & $A_{1g} \oplus 2A_{2g} \oplus B_{1g} \oplus B_{2g} \oplus 3E_g$\\
$6$ & $2A_{1u} \oplus A_{2u} \oplus 2B_{1u} \oplus 2 B_{2u} \oplus 3 E_u$\\
\bottomrule
\end{tabular}
%\end{small}
\end{minipage}
\caption{Transformation properties of the intrinsic electric (left) \cite{GTW95} and magnetic multipole operators (right) under the elements of the $\mathcal{D}_{4h}$ group. The results for the $2\lambda+1$ spherical components are grouped by their multipolarity $\lambda$.}
\label{tab:transprop_intrinsic_EM_operators}
\end{table}

For the electric multipole transitions with $\lambda\geq 1$, the operator can be written as \cite{GrM96}
\begin{equation}
\Omega_{\lambda \mu}(E)= \int\mathrm{d}^3r~r^{\lambda}Y_{\lambda}^{\mu}(\theta,\phi)\rho(\mathbf{r})~,
\end{equation}
where $\rho(\mathbf{r})$ is the charge density, which, in the pointlike approximation for the $\alpha$-particles, becomes, 
\begin{equation}
\rho(\mathbf{r}) = 2e\sum_{i=1}^6\delta(\mathbf{r}- \mathbf{R}_i)~,
\end{equation}
where $\mathbf{R}_i = (\xi_i,\eta_i,\zeta_i)$ are the coordinates of the clusters in the body-fixed frame. 
In particular, the $\Omega_{\lambda\nu}(E)$ operator transforms according to the irreducible representation $A_{1g}$ ($A_{1u}$) of $\mathcal{D}_{4h}$ if $\lambda$ is even (odd), irrespective on the spherical component $\nu$.
However, in the body-fixed frame, the components $\omega_{\lambda\nu}(E)$ have different transformation properties under the $\mathcal{D}_{4h}$ group \cite{GTW95}, summarized for all the spherical components with a given multipolarity $\lambda$ in Tab.~\ref{tab:transprop_intrinsic_EM_operators}.

Furthermore, the magnetic multipole operators ($\lambda \geq 1$) take the form 
\begin{equation}
\Omega_{\lambda \mu}(M)= \int\mathrm{d}^3r~\mathbf{j}(\mathbf{r})\cdot \mathbf{J}~r^{\lambda}Y_{\lambda}^{\mu}(\theta,\phi)~,
\end{equation}
where $\mathbf{J} \equiv (J_{\xi},J_{\eta},J_{\zeta})$ is the angular momentum operator in the laboratory frame \cite{Ste15,Ste26} and $\mathbf{j}(\mathbf{r})$ is the current density, which, in the pointlike approximation for the \ce{^{4}He} clusters, becomes 
\begin{equation}
\mathbf{j}(\mathbf{r}) = \frac{2e\hbar}{2mi}\sum_{i=1}^Z[\delta(\mathbf{r}-\mathbf{R}_i)\overrightarrow{\nabla}_i-\overleftarrow{\nabla}_i  \delta(\mathbf{r}-\mathbf{R}_i)]~,
\end{equation}
where $\overrightarrow{\nabla}_i$ refers to the $\alpha$-cluster with coordinate $\mathbf{R}_i$ in the laboratory frame.
By exploiting the Wigner-Eckart theorem \cite{GrM96}, the matrix elements on the r.h.s. of Eq.~\eqref{eqn:reducedEMtransitionprobabibilites} can be separated into rotational and vibrational matrix elements. When the LO rotational-vibrational states in Eq.~\eqref{eqn:rovibrationalstates_LO_D4h} are not factorized, rotational and vibrational matrix elements can be still separated at LO in the G$\alpha$CM expansion. 

With reference to Eq.~\eqref{eqn:reducedEMtransitionprobabibilites}, the overall matrix elements between the $H_{LO}$ eigenstates are subject to the customary parity rules,
\begin{subequations}
    \begin{equation}
        \pi_f = (-1)^{\lambda}\pi_i~,
\label{eqn:parityrule_Etrans}
    \end{equation}
    \begin{equation}
        \pi_f = (-1)^{\lambda+1}\pi_i~,
        \label{eqn:parityrule_Mtrans}
    \end{equation}
\end{subequations}
for electric and magnetic $2^{\lambda}$-pole transitions respectively. Eqs.~\eqref{eqn:parityrule_Mtrans} and \eqref{eqn:parityrule_Mtrans} are merely a consequence of the \textit{vanishing integral rule} \cite{BuJ04} applied to the rotational-vibrational states,   
\begin{equation}
    \Gamma[(\Psi_{RV}^{\mathcal{D}_{4h}})_f] \otimes \Gamma[\Omega_{\lambda\nu}(R)] \otimes   \Gamma[(\Psi_{RV}^{\mathcal{D}_{4h}})_i]  \supset A_{1g}~.
    \label{vanishingIntrule_RVstates}
\end{equation}
The latter states that the trivial representation must be included in the reducible representation of $\mathcal{D}_{4h}$ according to which the transition matrix element $\langle (\Psi_{RV}^{\mathcal{D}_{4h}})_f|\Omega_{\lambda\nu}(R)|(\Psi_{RV}^{\mathcal{D}_{4h}})_i\rangle$ transforms. Additionally, the triangular inequality applied to the angular momentum of the initial and final states, together with the multipolarity of the transition operator adds the known constraints over $J_f$, $J_i$ and $\lambda$.

The presence of a point-group symmetry in the equilibrium $\alpha$-cluster configuration of the nucleus yields additional selection rules, stemming from the intrinsic degrees of freedom, \textit{i.e.} the normal coordinates. 
The application of the vanishing integral rule to the matrix elements between the vibrational states and the intrinsic transition operators, 
\begin{equation}
    \Gamma[(\psi_{V})_f] \otimes \Gamma[\omega_{\lambda\nu}(R)] \otimes   \Gamma[(\psi_{V})_i]  \supset A_{1g}~,
    \label{vanishingIntrule_Vstates}
\end{equation}
provides new restrictions. When $(\psi_{V})_i$ or $(\psi_{V})_f$ is replaced by $(\bar{\psi}_{V})_i$ or $(\bar{\psi}_{V})_f$, the rule in Eq.~\eqref{vanishingIntrule_Vstates} retains its validity. 

Concerning intraband transitions, the selection rules obtained from Eq.~\eqref{vanishingIntrule_Vstates} are presented in Tab.~\ref{tab:intrabandsVSelectionRules} for all the non-excited and singly excited rotational bands. For parity reasons, electric (magnetic) transitions have even (odd) multipolarity.

\begin{table}[htb!]
\centering
\begin{tabular}{ccc}
\toprule
$\Gamma[\psi_V]$ $\leftrightarrow$ $\Gamma[\psi_V]$ & \textsc{Electric} & \textsc{Magnetic}\\
\midrule
$A_{1g} (g.s.)$ $\leftrightarrow$ $A_{1g}(g.s.)$ & E2, E4, E6, \ldots & M5, \ldots\\
$A_{1g} (\omega_1)$ $\leftrightarrow$ $A_{1g}(\omega_1)$ & E2, E4, E6, \ldots & M5, \ldots\\
$A_{1g} (\omega_2)$ $\leftrightarrow$ $A_{1g}(\omega_2)$ & E2, E4, E6, \ldots & M5, \ldots\\
$A_{2u} (\omega_3)$ $\leftrightarrow$ $A_{2u}(\omega_3)$ & E2, E4, E6 \ldots & M5, \ldots\\
$B_{1g} (\omega_4)$ $\leftrightarrow$ $B_{1g}(\omega_4)$ & E2, E4, E6 \ldots & M5, \ldots\\
$B_{2g} (\omega_5)$ $\leftrightarrow$ $B_{2g}(\omega_5)$ & E2, E4, E6, \ldots & M5, \ldots\\
$B_{2u} (\omega_6)$ $\leftrightarrow$ $B_{2u}(\omega_6)$ & E2, E4, E6 \ldots & M5, \ldots\\
$E_{g} (\omega_7)$ $\leftrightarrow$ $E_{g}(\omega_7)$ & E2, E4, E6, \ldots & M1, M3, M5 \ldots\\
$E_{u} (\omega_8)$ $\leftrightarrow$ $E_{u}(\omega_8)$ & E2, E4, E6, \ldots & M1, M3, M5 \ldots\\
$E_{u} (\omega_9)$ $\leftrightarrow$ $E_{u}(\omega_9)$ & E2, E4, E6, \ldots & M1, M3, M5 \ldots\\
$E_g \otimes E_{u} (\omega_7,\omega_9)$ $\leftrightarrow$ $E_g \otimes E_{u} (\omega_7,\omega_9)$ & E0, E2, E4, E6, \ldots & M1, M3, M5 \ldots\\
\bottomrule
\end{tabular}
\caption{Selection rules for intraband transitions, internal to all the unexcited and singly-excited rotational bands, plus the $E_g \otimes E_{u} = A_{1u} \oplus A_{2u} \oplus B_{1u} \oplus B_{2u}$ band with $\mathfrak{n}_7=1$ and $\mathfrak{n}_9=1$.}
\label{tab:intrabandsVSelectionRules}
\end{table}

Furthermore, since in non-excited and singly-excited bands no states with the same angular momentum are repeated in the same $|K|^{\pi}$-band, E0 transitions do not occur. Additionally, in non-degenerate bands, intraband M1 and M3 transitions are absent (cf. Tab.~\eqref{vanishingIntrule_Vstates}). Among states composing doubly-excited bands, such as the hypothetical $E_g\otimes E_u$ band with $\mathfrak{n}_7=1$ and $\mathfrak{n}_9=1$ based on the $0_1^-$ state discussed in the previous section, E0 transitions could take place, since two states with the same angular momentum are allowed to coexist in the same $|K|^{\pi}$ band.

\begin{table}[htb!]
\centering
\begin{tabular}{ccc}
\toprule
$\Gamma[\psi_V]$ $\leftrightarrow$ $\Gamma[\psi_V]$ & \textsc{Electric} & \textsc{Magnetic}\\
\midrule
$A_{1g} (g.s.)$ $\leftrightarrow$ $A_{1g}(\omega_1)$ & E0, E2, E4, E6, \ldots & M5, \ldots\\
$A_{1g} (g.s.)$ $\leftrightarrow$ $A_{1g}(\omega_2)$ & E0, E2, E4, E6, \ldots & M5, \ldots\\
$A_{1g} (g.s.)$ $\leftrightarrow$ $A_{2u}(\omega_3)$ & E3, E5, \ldots & M4, M6, \ldots\\
$A_{1g} (g.s.)$ $\leftrightarrow$ $B_{1g}(\omega_4)$ & E2, E4, E6 \ldots & M3, M5, \ldots\\
$A_{1g} (g.s.)$ $\leftrightarrow$ $B_{2g}(\omega_5)$ & E2, E4, E6 \ldots & M3, M5, \ldots\\
$A_{1g} (g.s.)$ $\leftrightarrow$ $B_{2u}(\omega_6)$ & E3, E5, \ldots & M2, M4, M6, \ldots\\
$A_{1g} (g.s.)$ $\leftrightarrow$ $E_{g}(\omega_7)$ & E2, E4, E6 \ldots & M1, M3, M5, \ldots\\
$A_{1g} (g.s.)$ $\leftrightarrow$ $E_{u}(\omega_8)$ & E3, E5, \ldots & M2, M4, M6 \ldots\\
$A_{1g} (g.s.)$ $\leftrightarrow$ $E_{u}(\omega_9)$ & E3, E5, \ldots & M2, M4, M6 \ldots\\
\bottomrule
\end{tabular}
\caption{Selection rules between non-excited and singly-excited rotational bands.}
\label{tab:groundstateband-singexcbandsVSelectionRules}
\end{table}

Concerning interband transitions, the application of the vanishing integral rule in Eq.~\eqref{vanishingIntrule_Vstates} delivers the results in Tabs.~\ref{tab:groundstateband-singexcbandsVSelectionRules} and \ref{tab:singexcbands-singexcbandsVSelectionRules} for the transitions between states with at most a single quantum of vibrational excitation.  

\begin{table}[htb!]
\centering
\begin{tabular}{ccc}
\toprule
$\Gamma[\psi_V]$ $\leftrightarrow$ $\Gamma[\psi_V]$ & \textsc{Electric} & \textsc{Magnetic}\\
\midrule
$A_{1g} (\omega_1)$ $\leftrightarrow$ $A_{1g}(\omega_2)$ & E0, E2, E4, E6, \ldots & M5, \ldots\\
$A_{1g} (\omega_1)$ $\leftrightarrow$ $A_{2u}(\omega_3)$ & E3, E5, \ldots & M4, M6, \ldots\\
$A_{1g} (\omega_1)$ $\leftrightarrow$ $B_{1g}(\omega_4)$ & E2, E4, E6, \ldots & M3, M5, \ldots\\
$A_{1g} (\omega_1)$ $\leftrightarrow$ $B_{2g}(\omega_5)$ & E2, E4, E6, \ldots & M3, M5, \ldots\\
$A_{1g} (\omega_1)$ $\leftrightarrow$ $B_{2u}(\omega_6)$ & E3, E5, \ldots & M2, M4, M6, \ldots\\
$A_{1g} (\omega_1)$ $\leftrightarrow$ $E_{g}(\omega_7)$ & E2, E4, E6, \ldots & M1, M3, M5, \ldots\\
$A_{1g} (\omega_1)$ $\leftrightarrow$ $E_{u}(\omega_8)$ & E3, E5, \ldots & M2, M4, M6 \ldots\\
$A_{1g} (\omega_1)$ $\leftrightarrow$ $E_{u}(\omega_9)$ & E3, E5, \ldots & M2, M4, M6 \ldots\\
$A_{1g} (\omega_2)$ $\leftrightarrow$ $A_{2u}(\omega_3)$ & E3, E5, \ldots & M4, M6, \ldots\\
$A_{1g} (\omega_2)$ $\leftrightarrow$ $B_{1g}(\omega_4)$ & E2, E4, E6, \ldots & M3, M5, \ldots\\
$A_{1g} (\omega_2)$ $\leftrightarrow$ $B_{2g}(\omega_5)$ & E2, E4, E6, \ldots & M3, M5, \ldots\\
$A_{1g} (\omega_2)$ $\leftrightarrow$ $B_{2u}(\omega_6)$ & E3, E5, \ldots & M2, M4, M6, \ldots\\
$A_{1g} (\omega_2)$ $\leftrightarrow$ $E_{g}(\omega_7)$ & E2, E4, E6, \ldots & M1, M3, M5, \ldots\\
$A_{1g} (\omega_2)$ $\leftrightarrow$ $E_{u}(\omega_8)$ & E3, E5, \ldots & M2, M4, M6 \ldots\\
$A_{1g} (\omega_2)$ $\leftrightarrow$ $E_{u}(\omega_9)$ & E3, E5, \ldots & M2, M4, M6 \ldots\\
$A_{2u} (\omega_3)$ $\leftrightarrow$ $B_{1g}(\omega_4)$ & E3, E5 \ldots & M2, M4, M6, \ldots\\
$A_{2u} (\omega_3)$ $\leftrightarrow$ $B_{2g}(\omega_5)$ & E3, E5 \ldots & M2, M4, M6, \ldots\\
$A_{2u} (\omega_3)$ $\leftrightarrow$ $B_{2u}(\omega_6)$ & E2, E4, E6, \ldots & M3, M5, \ldots\\
$A_{2u} (\omega_3)$ $\leftrightarrow$ $E_{g}(\omega_7)$ & E3, E5, \ldots & M2, M4, M6, \ldots\\
$A_{2u} (\omega_3)$ $\leftrightarrow$ $E_{u}(\omega_8)$ & E2, E4, E6, \ldots & M1, M3, M5, \ldots\\
$A_{2u} (\omega_3)$ $\leftrightarrow$ $E_{u}(\omega_9)$ & E2, E4, E6, \ldots & M1, M3, M5, \ldots\\
$B_{1g} (\omega_4)$ $\leftrightarrow$ $B_{2g}(\omega_5)$ & E4, E6, \ldots & M1, M3, M5, \ldots\\
$B_{1g} (\omega_4)$ $\leftrightarrow$ $B_{2u}(\omega_6)$ & E3, E5, \ldots & M4, M6, \ldots\\
$B_{1g} (\omega_4)$ $\leftrightarrow$ $E_{g}(\omega_7)$ & E2, E4, E6, \ldots & M1, M3, M5, \ldots\\
$B_{1g} (\omega_4)$ $\leftrightarrow$ $E_{u}(\omega_8)$ & E3, E5, \ldots & M2, M4, M6, \ldots\\
$B_{1g} (\omega_4)$ $\leftrightarrow$ $E_{u}(\omega_9)$ & E3, E5, \ldots & M2, M4, M6, \ldots\\
$B_{2g} (\omega_5)$ $\leftrightarrow$ $B_{2u}(\omega_6)$ & E3, E5, \ldots & M4, M6, \ldots\\
$B_{2g} (\omega_5)$ $\leftrightarrow$ $E_{g}(\omega_7)$ & E2, E4, E6, \ldots & M1, M3, M5, \ldots\\
$B_{2g} (\omega_5)$ $\leftrightarrow$ $E_{u}(\omega_8)$ & E3, E5, \ldots & M2, M4, M6, \ldots\\
$B_{2g} (\omega_5)$ $\leftrightarrow$ $E_{u}(\omega_9)$ & E3, E5, \ldots & M2, M4, M6, \ldots\\
$B_{2u} (\omega_6)$ $\leftrightarrow$ $E_{g}(\omega_7)$ & E3, E5, \ldots & M2, M4, M6, \ldots\\
$B_{2u} (\omega_6)$ $\leftrightarrow$ $E_{u}(\omega_8)$ & E2, E4, E6, \ldots & M1, M3, M5, \ldots\\
$B_{2u} (\omega_6)$ $\leftrightarrow$ $E_{u}(\omega_9)$ & E2, E4, E6, \ldots & M1, M3, M5, \ldots\\
$E_{g} (\omega_7)$ $\leftrightarrow$ $E_{u}(\omega_8)$ & E3, E5, \ldots & M2, M4, M6, \ldots\\
$E_{g} (\omega_7)$ $\leftrightarrow$ $E_{u}(\omega_9)$ & E3, E5 \ldots & M2, M4, M6, \ldots\\
$E_{u} (\omega_8)$ $\leftrightarrow$ $E_{u}(\omega_9)$ & E0, E2, E4, E6, \ldots & M1, M3, M5, \ldots\\
\bottomrule
\end{tabular}
\caption{Selection rules between singly-excited rotational bands.}
\label{tab:singexcbands-singexcbandsVSelectionRules}
\end{table}

Furthermore, when odd electric multipolarities are allowed, the E1 mode must be ruled out. In fact, all the transitions of E1 type vanish, since in the G$\alpha$CM protons and neutrons cannot move in opposition of phase, as they are bound within $\alpha$-clusters. Consequently, all the measured E1 transitions in the low-lying spectrum of \ce{^{24}Mg} entail the excitation of degrees of freedom internal to the $\alpha$-clusters.

\begin{table}[htb!]
\centering
\begin{tabular}{ccc}
\toprule
$\Gamma[\psi_V]$ $\leftrightarrow$ $\Gamma[\psi_V]$ & \textsc{Electric} & \textsc{Magnetic}\\
\midrule
$A_{1u} \oplus A_{2u} \oplus B_{1u} \oplus B_{2u} (\omega_7,\omega_9)$ $\leftrightarrow$ $A_{1g}(g.s.)$ & E3, E5, \ldots  & M2, M4, M6, \ldots\\
$A_{1u} \oplus A_{2u} \oplus B_{1u} \oplus B_{2u} (\omega_7,\omega_9)$ $\leftrightarrow$ $A_{1g}(\omega_1)$ & E3, E5, \ldots  & M2, M4, M6, \ldots\\
$A_{1u} \oplus A_{2u} \oplus B_{1u} \oplus B_{2u} (\omega_7,\omega_9)$ $\leftrightarrow$ $A_{1g}(\omega_2)$ & E3, E5, \ldots  & M2, M4, M6, \ldots\\
$A_{1u} \oplus A_{2u} \oplus B_{1u} \oplus B_{2u} (\omega_7,\omega_9)$ $\leftrightarrow$ $A_{2u}(\omega_3)$ & E0, E2, E4, E6, \ldots & M1, M3, M5, \ldots\\
$A_{1u} \oplus A_{2u} \oplus B_{1u} \oplus B_{2u} (\omega_7,\omega_9)$ $\leftrightarrow$ $B_{1g}(\omega_4)$ & E3, E5, \ldots & M2, M4, M6, \ldots\\
$A_{1u} \oplus A_{2u} \oplus B_{1u} \oplus B_{2u} (\omega_7,\omega_9)$ $\leftrightarrow$ $B_{2g}(\omega_5)$ & E3, E5, \ldots  & M2, M4, M6, \ldots\\
$A_{1u} \oplus A_{2u} \oplus B_{1u} \oplus B_{2u} (\omega_7,\omega_9)$ $\leftrightarrow$ $B_{2u}(\omega_6)$ & E0, E2, E4, E6, \ldots  & M1, M3, M5, \ldots\\
$A_{1u} \oplus A_{2u} \oplus B_{1u} \oplus B_{2u} (\omega_7,\omega_9)$ $\leftrightarrow$ $E_{g}(\omega_7)$ & E3, E5, \ldots & M2, M4, M6, \ldots\\
$A_{1u} \oplus A_{2u} \oplus B_{1u} \oplus B_{2u} (\omega_7,\omega_9)$ $\leftrightarrow$ $E_{u}(\omega_8)$ & E2, E4, E6, \ldots & M1, M3, M5, \ldots\\
$A_{1u} \oplus A_{2u} \oplus B_{1u} \oplus B_{2u} (\omega_7,\omega_9)$ $\leftrightarrow$ $E_{u}(\omega_9)$ & E2, E4, E6, \ldots & M1, M3, M5, \ldots\\
\bottomrule
\end{tabular}
\caption{Selection rules between the doubly-excited band $E_g \otimes E_{u} = A_{1u} \oplus A_{2u} \oplus B_{1u} \oplus B_{2u}$ band with $\mathfrak{n}_7=1$ and $\mathfrak{n}_9=1$ excitation quanta, and the non-excited or the singly-excited bands.}
\label{tab:nonexcited&singexcbands-doublyexcbandVSelectionRules}
\end{table}

From Tabs.~\ref{tab:groundstateband-singexcbandsVSelectionRules} and \ref{tab:singexcbands-singexcbandsVSelectionRules}, it follows that transitions involving rotational bands of $A_{1g}$ type do not allow M3 transitions. Additionally, transitions between $A_{1g}$ or $A_{2u}$ bands and $B_{1g}$, $B_{1u}$, $B_{2u}$, $E_g$ or $E_u$ bands do not include the ones of E0 type. In contrast, transitions between bands with excited doubly-degenerate modes do not receive any further restriction than the ones imposed by parity.
Unlike \ce{^{12}C} in the $\alpha$-cluster model with $\mathcal{D}_{3h}$ symmetry \cite{SFV16,FSV17}, for \ce{^{24}Mg} M1 transitions are present also between singly-excited rotational bands, although less frequently than $E2$ transitions. 
A series of interband reduced transition probabilities between states assigned to singly-excited rotational bands have been calculated at LO in the G$\alpha$CM framework \cite{Ste26}, and display an appreciable agreement with the experimental counterparts. 

Eventually, regarding the possible doubly excited band with $\mathfrak{n}_7=1$ and $\mathfrak{n}_9=1$, the selection rules with respect to transitions to non-excited and singly-excited rotational bands have been presented in Tab.~\ref{tab:nonexcited&singexcbands-doublyexcbandVSelectionRules}. For the latter, the restrictions imposed by $\mathcal{D}_{4h}$ symmetry through the vanishing integral rule are absent.

\section{Conclusion}

Motivated by the recent study on \ce{^{20}Ne} \cite{BiI21-01}, the $\alpha$-cluster structure of \ce{^{24}Mg} has been investigated, with special regard to the electromagnetic transitions. On the occasion, an approximation scheme for the implementation of the coupling between rotational and vibrational motion, based on the Watson Hamiltonian \cite{Wat68}, has been introduced. The formal developments of the G$\alpha$CM are detailed in Ref.~\cite{Ste26}.

The square bipyramid with $\mathcal{D}_{4h}$ symmetry, presented in Ref.~\cite{HaD66}, has been reproposed as an equilibrium $\alpha$-cluster configuration for \ce{^{24}Mg}. Thanks to the significant amount of spectroscopic data recorded since then, all the $9$ predicted singly-excited rotational bands have been identified \cite{Ste26}. In the analysis of the vibrations of the $6\alpha$-structure, the connection between certain normal modes ($A_{1g}$, $A_{2u}$, $E_g$) and low-energy cluster-decay channels has been highlighted.
Further support to the $\mathcal{D}_{4h}$-symmetric \textit{ansatz} has been provided by the calculated intraband and interband reduced electric and magnetic multipole transition probabilities between the identified $\alpha$-cluster states of \ce{^{24}Mg} in Ref.~\cite{Ste26}. 

The perturbative application of the G$\alpha$CM at NLO \cite{Ste26} is envisaged, since the rotation-vibration coupling is responsible of the changes in the nuclear moments of inertia in the excited rotational bands. Additionally, the tentative inspection of doubly-excited bands, with special attention to neighbour states of decay thresholds at $13.93$ MeV (\ce{^{12}C}$+$\ce{^{12}C}), 14.05 MeV ($2\alpha$+\ce{^{16}O}) and 21.21 MeV ($3\alpha+$\ce{^{12}C}).

After focusing on the transformation properties of the rotational and vibrational states at LO under the elements of $\mathcal{D}_{4h}$ group, the selection rules stemming from the intrinsic degrees of freedom of the system have been analyzed in the same fashion as in Refs.~\cite{SFV16,FSV17}.

The latter, summarized in Tabs.~\ref{tab:intrabandsVSelectionRules}-\ref{tab:nonexcited&singexcbands-doublyexcbandVSelectionRules} possess the dual purpose of providing a criterion for the classification of further observed lines into excited rotational bands predicted by the G$\alpha$CM as well as of delivering guidance for the 
identification of violations of $\mathcal{D}_{4h}$ symmetry in the spectrum.

With the development of new facilities such as the \textit{variable energy gamma system} (VEGA) at ELI-NP (Măgurele, Romania) where photo-excitation experiments in the energy range $1$-$20$~MeV plays a major role, the full understanding of electromagnetic selection rules as a tool to assess the validity of nuclear structure models is mandatory \cite{FSV17}.

\section*{Acknowledgements}

This research is dedicated to the memory of Andrea Vitturi (University of Padova), co-supervisor of the M.Sc. thesis of G.S. in 2015 \cite{Ste15}, as well as coauthor of Refs.~\cite{SFV16,FSV17}.
This proceeding is part of a greater work started under the impulse of the co-author (31.07.1938 - 19.06.2023), who carried out the first calculations in the initial phase. Moreover, K.-H. S. was fascinated by the phenomenology of clustering in nuclei and, throughout his career, contributed to several experimental campaigns for the investigation of EM properties of light nuclei, namely magnetic dipole moments \cite{SSM22}.
G.S. acknowledges funding form the \emph{Espace de Structure et de réactions Nucléaires Théorique} (ESNT) of the CEA/DSM-DAM and expresses gratitude to Dean Lee (Michigan State University) and Vittorio Somà (CEA Paris-Saclay) for the regular discussions.

\end{document}